\documentclass[AMA,STIX2COL]{MRM}
\articletype{PREPRINT - SUBMITTED TO MAGNETIC RESONANCE IN MEDICINE}%

\received{August 6, 2026}
\revised{XXX}
\accepted{XXX}
\usepackage{amsmath}
\usepackage{amssymb}
\usepackage{booktabs}     
\usepackage{makecell}     
\usepackage{multirow}     
\usepackage{threeparttable} 
\usepackage{graphicx}
\usepackage{algorithm}
\usepackage{algpseudocode}
\usepackage{float}

\makeatletter
\Gin@draftfalse
\makeatother

\begin{document}

\title{Highly accelerated 3D Cartesian MPnRAGE with implicit neural representation reconstruction}

\author[1,2]{Natascha Niessen}{\orcid{0009-0004-2629-3638}}
\author[2,7]{Ana Beatriz Solana}{\orcid{0000-0002-7390-0101}}
\author[2]{Carolin M. Pirkl}{\orcid{0000-0002-5759-5290}}
\author[2,3]{Tim Sprenger}{\orcid{0000-0001-7037-8605}}
\author[1,4]{Hannah Eichhorn}{\orcid{0000-0001-6980-9703}}
\author[1,4]{Veronika Spieker}{\orcid{0000-0001-7720-7569}}
\author[1]{Wenqi Huang}{\orcid{0000-0002-0155-5409}}
\author[2]{Rolf F. Schulte}{\orcid{0000-0002-1334-1264}}
\author[2]{Florian Wiesinger}{\orcid{0000-0002-5597-6057}}
\author[7]{Tobias C. Wood}{\orcid{0000-0001-7640-5520}}
\author[2,5,6]{Marion I. Menzel}{\orcid{0000-0003-0087-9134}}
\author[1,4,8,9]{Julia A. Schnabel}{\orcid{0000-0001-6107-3009} on behalf of the PREDICTOM consortium}

\authormark{NATASCHA NIESSEN \textsc{et al}}

\address[1]{\orgdiv{School of Computation and Information Technology (CIT)}, \orgname{Technical University of Munich}, \orgaddress{\state{Munich}, \country{Germany}}}

\address[2]{\orgname{GE HealthCare}, \orgaddress{\state{Munich}, \country{Germany}}}

\address[3]{\orgdiv{Department of Clinical Neuroscience}, \orgname{Karolinska Institutet}, \orgaddress{\state{Stockholm}, \country{Sweden}}}

\address[4]{\orgdiv{Institute of Machine Learning in Biomedical Imaging}, \orgname{Helmholtz Munich}, \orgaddress{\state{Neuherberg}, \country{Germany}}}

\address[5]{\orgname{Technische Hochschule Ingolstadt}, \orgaddress{\state{Ingolstadt}, \country{Germany}}}

\address[6]{\orgdiv{School of Natural Sciences}, \orgname{Technical University of Munich}, \orgaddress{\state{Munich}, \country{Germany}}}

\address[7]{\orgdiv{Department of Neuroimaging}, \orgname{King's College London}, \orgaddress{\state{London}, \country{United Kingdom}}}

\address[8]{\orgdiv{School of Biomedical Engineering and Imaging Sciences}, \orgname{King's College London}, \orgaddress{\state{London}, \country{United Kingdom}}}

\address[9]{ \orgname{Munich Center for Machine Learning (MCML)}, \orgaddress{\state{Munich}, \country{Germany}}}

\corres{Natascha Niessen \\ \email{natascha.niessen@tum.de}}

\finfo{IHI JU No 101132356, UKRI No 10083467, KCL 10083181, Exeter No 10091560, Geneva SERI No 113152304, ERC No 884622}

\abstract[Abstract]{
\section{Purpose} MPnRAGE enables multiple inversion contrast images in a single scan, allowing quantitative T1 mapping, tissue nulled contrasts, and standard MPRAGE synthesis. However, current 3D scan times remain clinically impractical, motivating accelerated 3D MPnRAGE.
\section{Methods} This work provides a highly accelerated Cartesian 3D MPnRAGE sequence with joint implicit neural representation (INR) reconstruction. The sequence uses a tailored view-ordering strategy, flip angle schedule and complementary variable-density Poisson-disk undersampling. Calibration data acquired during otherwise unused delay time are used for sensitivity map estimation and complementary high-frequency sampling.
\section{Results} Ten INR-reconstructed inversion images at 1.5 mm$^3$ are evaluated against fully sampled references via retrospective undersampling. Prospectively accelerated 1 mm$^3$ images at R ~=~ 20 (5.39 min) demonstrate clinical feasibility. INR reconstruction outperforms subspace and iterative local low rank reconstruction on highly accelerated data.
\section{Conclusion}
The proposed highly undersampled 3D Cartesian MPnRAGE with INR reconstruction generates multiple high-quality inversion contrasts in substantially reduced scan time. Scan-specific INR reconstruction improves image quality while reducing reconstruction time versus state-of-the-art methods.
}



\wordcount{4900}

\jnlcitation{\cname{%
\author{N. Niessen},
\author{A.B. Solana},
\author{C.M. Pirkl},
\author{T. Sprenger},
\author{H. Eichhorn},
\author{V. Spieker},
\author{W. Huang},
\author{R.F. Schulte},
\author{F. Wiesinger},
\author{T.C. Wood},
\author{M.I. Menzel}, and
\author{J.A. Schnabel}, on behalf of the PREDICTOM consortium} (\cyear{2026}),
\ctitle{Highly accelerated 3D Cartesian MPnRAGE with implicit neural representation reconstruction}, \cjournal{Magn. Reson. Med.}, \cvol{year;vol:pages}.}

\maketitle

\section{Introduction}\label{sec1}

The Magnetization-Prepared Rapid Gradient-Echo (MPRAGE)\cite{mugler1990mprage} sequence has become widely used for clinical brain anatomical imaging as it can generate a single inversion recovery (IR) T1-weighted contrast comparatively quickly. MP2RAGE, an extension of MPRAGE, enables the acquisition of two inversion contrasts in a single scan, allowing for bias-field–robust T1 mapping alongside the synthesis of a conventional T1-weighted image and fluid and white matter suppression (FLAWS) images \cite{Marques2010_MP2RAGE} \cite{Tanner2012_FLAWSMP2RAGE}. MPnRAGE further extends this concept by acquiring data continuously throughout the IR, enabling quantitative T1 mapping and the generation of multiple contrasts from a single dataset\cite{Kecskemeti2016_MPnRAGE}. The clinical potential of MPnRAGE has been demonstrated, for example, in multiple sclerosis (MS), where enhanced lesion detection and longitudinal sensitivity of T1-based measures have been reported \cite{Kecskemeti2015_MS_MPnRAGE}. Furthermore, multiple IR GRE data have been identified as a potential biomarker for myelin state and axonal damage in MS lesions \cite{Gkotsoulias2026_ISMRM}.

Recent progress in deep learning–based MRI reconstruction has enabled substantial scan time reductions, including highly accelerated and multi-parametric acquisitions \cite{Hammernik2023, Heckel2021}. For MPnRAGE, the deep factor model based on convolutional neural networks (CNN) has been demonstrated to further accelerate the original radial sequence by exploiting shared contrast structure across inversion times (TI) \cite{Chen2025_RadialMPnRAGE}. Recently, implicit neural representation (INR) networks have emerged as a promising alternative for multiparametric MRI reconstruction \cite{feng2025refine,zhang2026lorein,lao2025summit}, representing images as continuous functions over spatial and contrast dimensions. INR-based methods are self-supervised and scan-specific, making them particularly attractive for new sequences such as MPnRAGE, where large training datasets are unavailable.

As MPnRAGE continuously samples the signal evolution during IR, it is particularly well suited to INR-based reconstruction, where continuous contrast variation can be modeled jointly with spatial structure. Early work already suggested that incorporating high spatial frequency components acquired across the IR curve may allow for even higher accelerations and improve SNR compared to single-contrast reconstructions \cite{Kecskemeti2016_MPnRAGE}. We recently introduced an INR framework for multi-contrast MRI that jointly embeds multiple TI images within a single network, using complementary Poisson-disk undersampling across contrasts to efficiently exploit k-space coverage \cite{ Niessen2025_MICCAI}, however based on retrospective undersampling.

Building on this framework, the present work implements complementary undersampling for Cartesian 3D MPnRAGE \cite{allen2026cartesian}, enabling prospective acceleration with a tailored view-ordering strategy that aims at limiting eddy current artifacts while maintaining TI continuity in k-space. The MPnRAGE sequence matches the inversion timing and recovery time (TR) of a conventional clinical MPRAGE, enabling direct generation of a standard MPRAGE-like T1-weighted image. In addition, the multi-contrast acquisition provides quantitative parameter maps as well as tissue-nulled contrasts, including gray-matter–nulled and white-matter–nulled images, from the same scan.

Additionally, the previously unused delay time between MPnRAGE acquisitions and the subsequent inversion is used to acquire calibration data with low flip angles (FA) and higher sampling density. Because this calibration data is acquired at very late TIs, when the longitudinal magnetization has already recovered to positive values, it can be used to obtain sensitivity maps directly from the MPnRAGE sequence while reducing sensitivity-map mismatch and minimizing phase errors associated with magnetization sign changes during recovery \cite{Kecskemeti2016_MPnRAGE}. Furthermore, this denser sampled sensitivity data is included in the joint INR reconstruction, improving the conditioning of the inverse problem. FA scheduling is used to counteract low SNR near tissue null-points along the IR curve. The proposed INR-based reconstruction is evaluated against state-of-the-art multi-contrast reconstruction methods.
In summary, we introduce an acquisition and reconstruction framework that integrates the following components:

\begin{enumerate}
    \item A Cartesian MPnRAGE sequence with a tailored view-ordering strategy for complementary variable-density Poisson-disk undersampling for highly accelerated 3D acquisitions.
    \item Acquisition of calibration data during previously unused delay time, increasing k-space coverage and enabling sensitivity map estimation.
    \item Flip-angle scheduling for improved SNR near tissue null points.
    \item Systematic evaluation of INR-based reconstruction for MPnRAGE against established reconstruction approaches.
\end{enumerate}

\section{Methods}\label{sec2}

\subsection{Accelerated Cartesian 3D MPnRAGE sequence}\label{cart_MPnRAGE_acqu}

The building blocks of the proposed accelerated Cartesian 3D MPnRAGE sequence are described in the following sections.

\subsubsection{Complementary undersampling with tailored view-ordering}\label{comp_Poisson}

The images resulting from MPnRAGE differ in contrast while sharing anatomical information. This redundancy in anatomical information across inversion contrast images can be leveraged to achieve high accelerations. The center of k-space containing crucial contrast information needs to be sampled densely while the higher frequencies can be sampled more sparsely and in a complementary fashion across contrasts.

\begin{figure*}[t]
\centerline{\includegraphics[width=\linewidth]{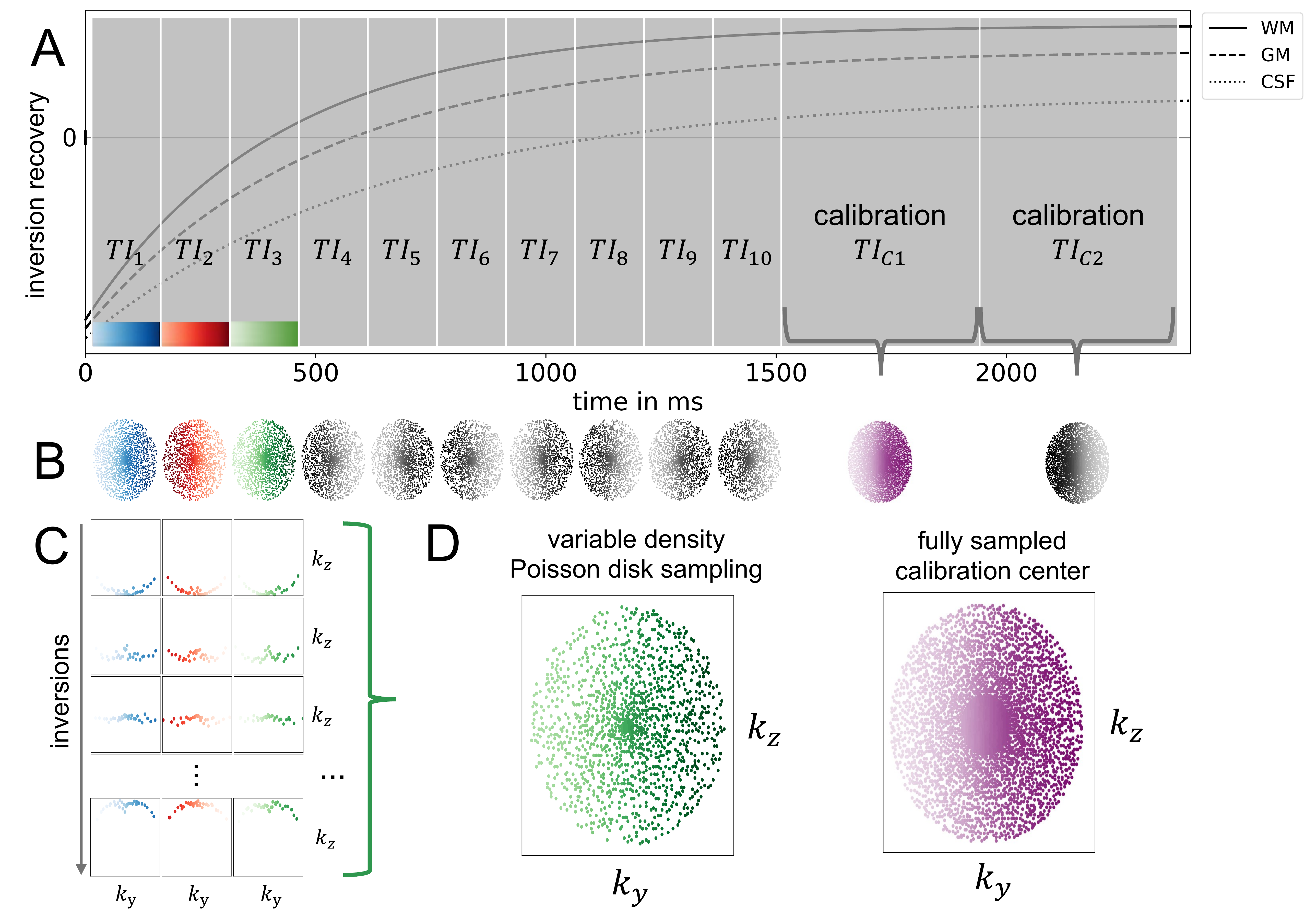}}
\caption{\hspace*{0.5em}
The proposed sampling strategy in the ky-kz plane (readout in kx): A) 10 regular TI volumes and two calibration volumes covering a larger time interval are acquired along the inversion recovery. B) k-space data are acquired with variable density Poisson-disk sampling patterns that are complementary across TIs. C) After every inversion pulse, a subset of kx lines is acquired for each TI bin. The acquisition order of the individual samples is indicated by the color shades (from light to dark colors). The view ordering was optimized for continuity in ky/kz with the k-space center traversal in the middle of the TI bin. Note that the k-space traversal direction in ky is alternated for every second TI (e.g. TI 1 sampling starts from the left, while TI 2 sampling starts from the right etc.). D) The union of all subsets acquired over all inversions for $TI_3$ at R ~=~ 20 and corresponding $TI_{C1}$. For calibration data, a larger fully sampled center is acquired e.g. for sensitivity map estimation.}\label{fig_sampling}
\end{figure*}

We use the strategy introduced by Levine et al. \cite{Levine2017_PoissonDisc}, in which the k-space of each individual TI is acquired using variable-density Poisson-disk sampling. To ensure complementarity across contrasts at higher frequencies, the different TI sampling patterns are designed so that their union also approximates Poisson-disk sampling (see Figure \ref{fig_sampling}B). The k-space center is densely sampled for each TI using a fully sampled central elliptical region with axis lengths of 10 x 6 samples, adapted to the anisotropic dimensions of the imaging matrix. Different fully sampled center sizes were evaluated under both retrospective and prospective undersampling, but their effect on reconstruction performance was limited.

We assess the impact of the complementarity of the Poisson-disk sampling across TI contrasts with retrospective undersampling and INR reconstruction for 1) Poisson-disk sampling where each TI contrast is sampled with the same sampling mask vs. 2) complementary Poisson-disk undersampling.

Figure \ref{fig_sampling} shows the data acquisition strategy where different inversion contrast images are acquired following an adiabatic inversion pulse. The acquisition time after the inversion pulse is divided into different TI bins. The undersampled k-spaces for the different TIs are acquired in an interleaved fashion: Following an inversion pulse, a subset of $P_{per TI bin}$ kx lines is acquired for each TI bin. This is repeated until all points of the targeted complementary undersampling masks have been acquired (Figure \ref{fig_sampling}C). 

The view ordering, i.e. the acquisition order of kx lines along the inversion recovery, is optimized for temporal and spatial continuity in the ky-kz plane to limit contrast blurring and reduce eddy current artifacts by minimizing abrupt phase-encoding gradient transitions \cite{Niessen2026_ISMRM}. This continuity is ensured within and across TI bins by acquiring one TI bin in the positive ky direction and the subsequent TI bin in the negative ky direction, alternating the direction for every other TI bin, as illustrated in Figure \ref{fig_sampling}C. The k-space center is acquired in the middle of the TI bin. $P_{per TI bin}$ was selected to match the timings of a clinically used T1 weighted MPRAGE sequence.

The 3D MPnRAGE sequence was implemented using the MNS Research Package (GE HealthCare, Chicago, USA), a flexible pulse sequence development platform.

\subsubsection{Integrated calibration TI for sensitivity maps}

In order to let the longitudinal magnetization recover before the next inversion, the clinically used T1 weighted MPRAGE sequence includes a recovery period before each inversion pulse. We use this delay time to acquire more densely sampled calibration data with a large fully sampled central elliptical k-space region of 40 x 32 samples, which can be used, for example, for sensitivity map estimation. The longer calibration TI intervals provide a longer readout window than the regular contrast TI, allowing to acquire more k-space points $P_{calibration}$. We choose to acquire two calibration intervals to limit the signal variation within one interval due to inversion recovery. A low FA of 2° for the calibration TI readouts ensures that the recovery of the longitudinal magnetization is not substantially disturbed. For generating sensitivity maps, the fully sampled center of k-space is Fermi-filtered.

\subsubsection{Iterative dictionary for T1 mapping and FA schedule optimization}\label{Iter_dict}

T1 mapping was performed using dictionary matching based on a signal model of the MPnRAGE acquisition. The model simulates the longitudinal magnetization evolution across repeated inversion cycles, accounting for inversion pulses, gradient-echo readouts with the applied flip-angle schedule, and T1 recovery. The simulation was iterated until a steady-state signal evolution was reached. The magnetization at the center of each TI interval, corresponding to the acquisition of the k-space center and thus the dominant image contrast, was used to generate the dictionary entries for T1 estimation.

\begin{figure}[t]
\centerline{\includegraphics[width=\linewidth]{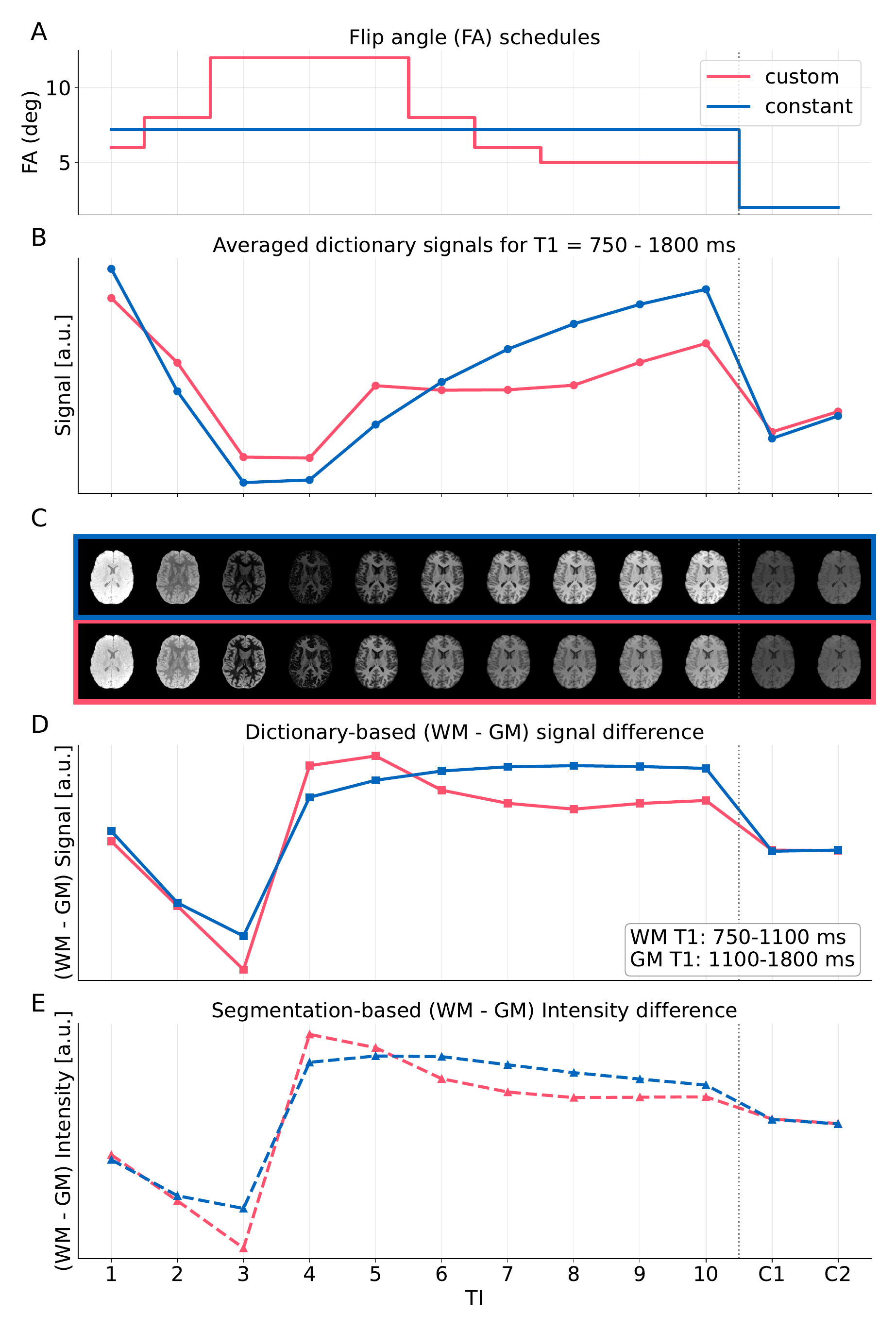}}
\caption{\hspace*{0.5em}
The effect of different FA schedules on signal intensity and white matter (WM) - gray matter (GM) contrast. A) Our proposed custom flip angle (FA) schedule (pink) and a constant FA schedule (blue). B) Dictionary signals averaged for T1 ~=~ 750 - 1800 ms. Signals for TI 2 - 5 with inherently low SNR are increased and TI 1, TI 6 - TI 10 are decreased with the custom FA schedule compared to the constant FA schedule. C) Reconstructed images showing stronger signal in inherently low SNR TI for the custom FA schedule. D) (WM - GM) signal difference calculated with the dictionary for WM T1 ~=~ 750 - 1100 ms and GM T1 ~=~ 1100 - 1800 ms. E) (WM - GM) intensity difference based on segmentation of images in C showing similar behavior as dictionary simulations from D. For calibration TI C1 and C2, FA is set to $2^\circ$ for both FA schedules in order to limit the destructive effect of the readout on the recovery of the longitudinal magnetization.
}
\label{fig_FA_schedule}
\end{figure}

As can be seen in \ref{fig_FA_schedule}A, the different tissues traverse their nulling points as they recover their longitudinal magnetization. Around this nulling point, the signal is very low in magnitude which results in a low signal-to-noise-ratio (SNR). In order to increase the SNR, we propose a custom FA schedule that is displayed in Figure \ref{fig_FA_schedule}A.
In order to determine the FA schedule, we calculated dictionary signals for a constant FA schedule and different custom FA schedules and averaged the dictionary signals for T1 values typical for white matter (WM) and gray matter (GM)\cite{ZavalaBojorquez2017}. The absolute values of the averaged signals are plotted in Figure~\ref{fig_FA_schedule}B. We determined the FA schedule as a tradeoff between increasing the signal of low SNR TIs (TI 2 - TI 5) and keeping the FA values in a reasonable range up to 12° to avoid saturation effects along the TI curve. For TI images that inherently provide high signal (TI 1, TI 6-TI 10), the FA is reduced in order to limit the destructive effect of the readout on the recovery of the longitudinal magnetization.

\subsection{Image reconstruction}

\subsubsection{Forward model}

For multi-contrast MRI reconstruction, the goal is to recover a set of images
$\mathbf{d} \in \mathbb{C}^{(V_y \times V_z)\times N}$ representing the same anatomy at $N$ different TIs. In this study, we focus on two-dimensional image slices of size $V_y \times V_z$, with $k_x$ denoting the readout dimension. Since the Fourier transform is separable across spatial axes, the readout FFT can be performed first, allowing the reconstruction to be split into independent smaller problems of size $(V_y \times V_z)\times N$ which improves the memory and computational efficiency of the INR reconstruction.

For each receive coil $c ~=~ 1,\ldots,C$, the acquired $k$-space measurements
$\mathbf{D}_c \in \mathbb{C}^{(V_y \times V_z)\times N}$ are modeled as

\begin{equation}
\mathbf{D}_c ~=~ \mathbf{M}\mathbf{F}\mathbf{S}_c\mathbf{d} + \mathbf{e}_c,
\end{equation}

\noindent where $\mathbf{F}$ denotes the two-dimensional Fourier encoding operator and
$\mathbf{S}_c$ is a diagonal matrix describing the spatial sensitivity of coil $c$. The term $\mathbf{e}_c$ accounts for additive white Gaussian noise in the measurements.

For the complementary undersampling framework (Section~\ref{comp_Poisson}), the sampling operator is written as $\mathbf{M}~~=~~[\mathbf{M}_1,\ldots,\mathbf{M}_N]$, where each TI contrast $n~~=~~1,\ldots,N$ is associated with a distinct undersampling mask $\mathbf{M}_n$.

\begin{figure*}[t]
\centerline{\includegraphics[width=\linewidth]{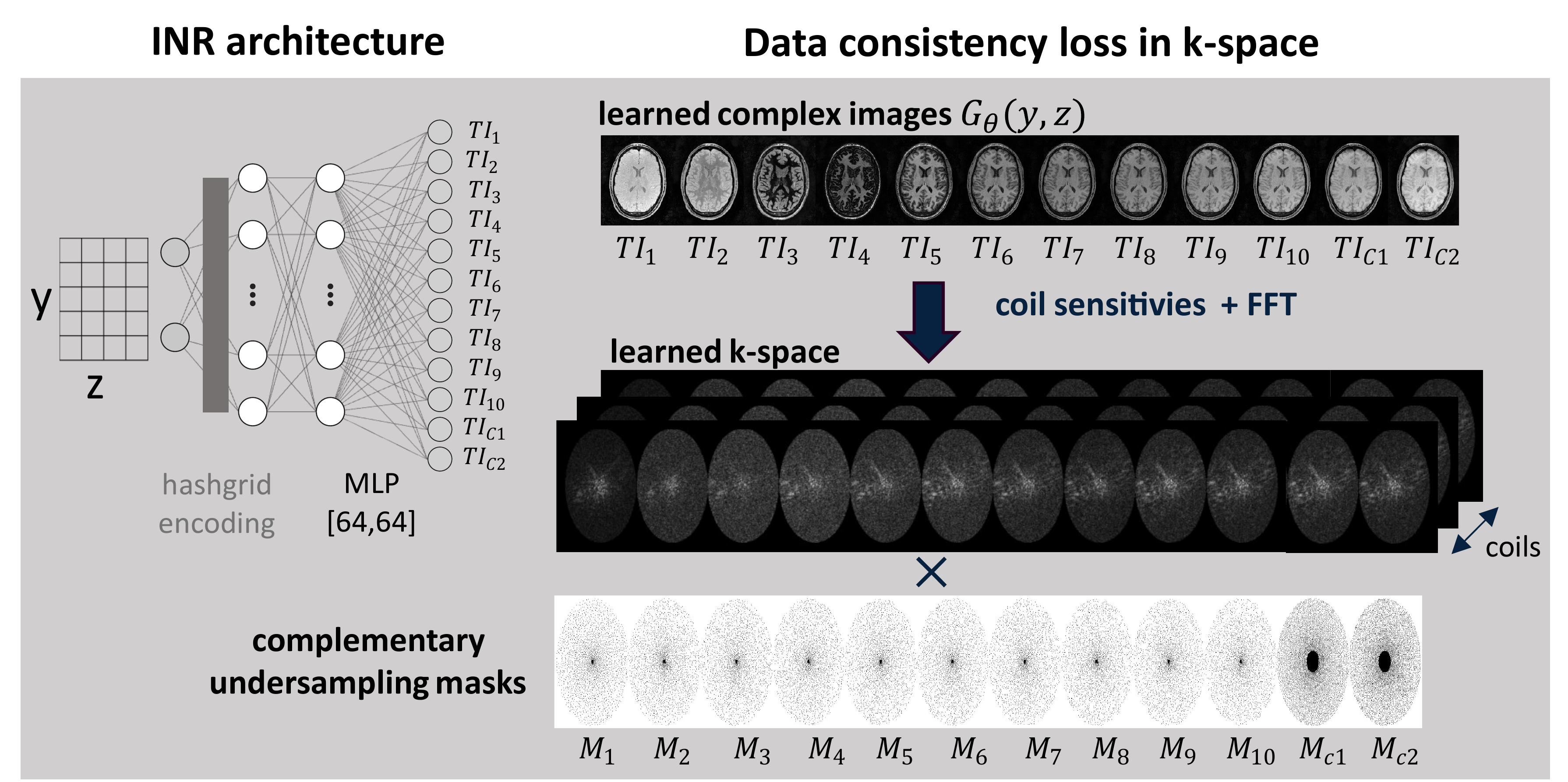}}
\caption{\hspace*{0.5em}Joint INR reconstruction with complementary undersampling. The INR is trained from scratch to represent the multi-TI images. At each iteration, the imaging forward operator with coil sensitivities, fast Fourier transform (FFT) and complementary sampling masks is applied to the learned images. The data consistency loss is calculated in k-space. Calibration TI are also used for training which enables to leverage further high frequency information from the previously unused delay time. The INR consists of hashgrid encoding and a lightweight MLP.}\label{fig_INR}
\end{figure*}

\subsubsection{Joint INR reconstruction}

The complementary Poisson-disk sampling places the k-space samples in an incoherent yet spatially balanced manner \cite{Levine2017_PoissonDisc} which results in noise-like aliasing artifacts that can effectively be mitigated by deep learning-based reconstructions \cite{Heckel2021,spieker2024deep}.
To enable reconstruction independent of a training dataset and leverage redundancies across the complementary sampled inversion contrasts, we use a multi-contrast INR network for reconstruction.
The INR learns a continuous mapping $G_\theta (\cdot): \mathbb{R}^2 \rightarrow \mathbb{C}^N$ from image coordinates $(y,z)\in \mathbb{R}^2$ to the complex voxel-wise signal evolution across $N$ contrasts with model parameters $\theta$. We jointly embed all contrasts within a single INR network leveraging the shared anatomical structure provided by complementary sampling.

At each iteration, the model outputs a representation $G_\theta (\cdot)$ of the images, referred to as ``learned images'' (Figure \ref{fig_INR}), to which we apply the forward model consisting of coil sensitivities and a fast Fourier transform (FFT) to obtain the ``learned k-space" for each contrast. Next, the complementary undersampling masks are applied to the learned k-space before comparing it to the acquired k-space data $D_c$ in the data consistency loss

\begin{equation}
\begin{gathered}
L(\theta) ~=~
\sum_{c~=~1}^{C}
\Big\|
W(\mathbf{k}_y,\mathbf{k}_z)
\cdot
\left[
\mathbf{M}\mathbf{F}\mathbf{S}_c G_\theta(\mathbf{y},\mathbf{z})
- \mathbf{D}_c
\right]
\Big\|_2^2
+ \lambda R(\theta) \\
\text{with} \quad
W(\mathbf{k}_y,\mathbf{k}_z)
~=~
\sqrt{\mathbf{k}_y^2 + \mathbf{k}_z^2} + 1 .
\end{gathered}
\end{equation}

\noindent using weights $W(\mathbf{k}_y,\mathbf{k}_z)$ proportional to the Euclidean distance from the k-space center as proposed in our previous work \cite{Niessen2025_MICCAI}. We use spatial total variation (TV) regularization $R(\theta) ~=~ \| \Delta G_\theta(\mathbf{y},\mathbf{z}) \|_1$ where $\Delta$ represents the gradient operator, weighted by $\lambda$. TV directly operates on the output of the INR corresponding to the reconstructed image $G_\theta(\mathbf{y},\mathbf{z})$ and eliminates image noise by enhancing local spatial consistency. $\lambda$ was determined via parameter sweep and kept constant for all datasets to show generalizability.

The INR architecture combines hashgrid encoding \cite{Mueller2022} and a lightweight multilayer perceptron (MLP) with two layers of 64 neurons each. It is trained from scratch for each subject and slice. We use the recommended parameters from the original hashgrid encoding implementation \cite{Mueller2022}: $L ~=~ 16$ resolution levels, 2 features per level and a base resolution of $N_{min}~=~16$. The hash-table size was determined as $2^{14}$ by a sweep ranging from $2^{12}$ to $2^{20}$. The level per scale $b$ was determined based on the largest spatial dimension $N_{max} ~=~ V_y$ as

\begin{equation}
b
~=~
\left(
\frac{N_{\max}}{N_{\min}}
\right)^{\frac{1}{L-1}}
\end{equation}

\subsection{Data acquisition}

Following the MPnRAGE acquisition scheme described in Section \ref{cart_MPnRAGE_acqu}, phantom and in-vivo images were acquired on a 3T DISCOVERY™ MR750w scanner using an 8-channel or 24-channel head coil and on a 3T SIGNA™ Premier scanner (both GE HealthCare, Chicago, USA) using a 48-channel head coil. The 48 channel and 24 channel data were compressed to 12 virtual coils via principal component analysis (PCA). 

For all images, a series of 10 TI contrast images followed by two calibration images were acquired with an adiabatic 180° inversion pulse and a series of spoiled gradient echo readouts for each inversion with a readout bandwidth of $\pm 31.25$ kHz. We conducted experiments for different resolutions using a FOV 24×24×19.2 cm³.

\subsubsection{Phantom data}

The quantitative NIST/ISMRM phantom \cite{Stupic2021_NISTPhantom} was imaged in sagittal orientation with the frequency encoding direction \makebox{anterior/posterior} to make the readout direction perpendicular to the T1 array plane of the phantom. All acquisitions in table \ref{tab1} were first tested on the phantom. Additionally, we acquired spin echo inversion recovery (SE-IR) data of the slice corresponding to vials of different T1 values to assess potential changes in quantitative T1 values with respect to the values in the documentation. This data acquisition took 13.36 hours.

\subsubsection{In vivo data}

In vivo data were acquired from five healthy subjects (26-42 years of age). The IRB was reviewed by the ``Ethikkommission der Bayerischen Landesärztekammer" (BLAEK) under the registration Nr. DE/EKBY10/00064163 and the Ethics Committee of the Faculty of Medicine, LMU Munich (project number 24-0701).

The scan plane was sagittal with the frequency encoding direction superior/inferior. At an isotropic resolution of $1.5\times1.5\times1.5~\mathrm{mm}^3$, fully sampled data were acquired from two healthy volunteers in 29.22 minutes and reconstructed with an iFFT to serve as reference images and for retrospective undersampling experiments. At the same resolution, prospectively accelerated acquisitions were acquired with acceleration factors $R~=~14$ (2.24 min), $R~=~17$ (2.00 min), and $R~=~20$ (1.45 min).

At an isotropic resolution of $1\times1\times1~\mathrm{mm}^3$, accelerated acquisitions were acquired from five healthy volunteers with acceleration factors $R~=~14$ (7.32 min), $R~=~17$ (6.29 min), and $R~=~20$ (5.39 min). Acquisition parameters and scan times are reported in Table~\ref{tab1}.

For all subjects we additionally acquired a product MPRAGE sequence for comparison with the following parameters: TI ~=~ 1000 ms, delay time 900 ms, FA 8° FOV 25.6 × 25.6 × 25.6 cm³, 1 $\mathrm{mm}^3$ resolution, readout bandwidth $\pm 31.25$ kHz and scan time 4.26 minutes. We optimized the parameters of our MPnRAGE sequence to closely match the timings of this MPRAGE sequence. The MPRAGE data was reconstructed using a product reconstruction.

\setlength{\tabcolsep}{2pt} 
\begin{table*}[t]
\caption{\hspace*{0.5em}MPnRAGE Acquisition Parameters and Times.\label{tab1}}
\centering
\begin{tabular*}{\textwidth}{@{\extracolsep{\fill}}cccccccccc@{}}
\toprule
\multirow{2}{*}{\makecell{\textbf{Resolution}\\\textbf{[$\mathrm{mm}^3$]}}}
& \textbf{Imaging}
& \textbf{MPnRAGE}
& \textbf{$\mathbf{TI_1}$}
& \textbf{$\mathbf{\Delta TI}$}
& \textbf{$\mathbf{\Delta TI_{cal}}$}
& \multirow{2}{*}{\makecell{\textbf{$\mathbf{P}_{\mathrm{per\,TI\,bin}}$}}}
& \multirow{2}{*}{\makecell{\textbf{$P_{\mathrm{calibration}}$}}}
& \textbf{Acceleration}
& \textbf{Acquisition} \\
&
\textbf{TR [ms]}
& \textbf{TR [s]}
& \textbf{[ms]}
& \textbf{[ms]}
& \textbf{[ms]}
&
&
& $\mathbf{R}$
& \textbf{Time [min]} \\
\midrule
\multirow{4}{*}{\centering $1.5\times1.5\times1.5$}
 & \multirow{4}{*}{\centering 6.8}
 & \multirow{4}{*}{\centering 2.44}
 & \multirow{4}{*}{\centering 74}
 & \multirow{4}{*}{\centering 151.8}
 & \multirow{4}{*}{\centering 445.0}
 & \multirow{4}{*}{\centering 22}
 & \multirow{4}{*}{\centering 66}
 & fully sampled & 29.22 \\
 & & & & & & & & 14 & 2.24 \\
 & & & & & & & & 17 & 2.00 \\
 & & & & & & & & 20 & 1.45 \\
\addlinespace[4pt]
\multirow{3}{*}{\centering $1\times1\times1$}
 & \multirow{3}{*}{\centering 8.9}
 & \multirow{3}{*}{\centering 2.43}
 & \multirow{3}{*}{\centering 76.2}
 & \multirow{3}{*}{\centering 151.3}
 & \multirow{3}{*}{\centering 455.4}
 & \multirow{3}{*}{\centering 17}
 & \multirow{3}{*}{\centering 50}
 & 14 & 7.32 \\
 & & & & & & & & 17 & 6.29 \\
 & & & & & & & & 20 & 5.39 \\
\bottomrule
\end{tabular*}
\begin{tablenotes}
\item
\end{tablenotes}
\end{table*}

\subsection{Comparison methods}

We compare our proposed INR reconstruction to zero-filled iFFT and parallel imaging compressed sensing (PICS) which both reconstruct each TI contrast image individually. Additionally, we compare to a subspace method and iterative local low rank (LLR) which both also reconstruct all contrasts jointly, leveraging the complementary acquired information.

For all reconstruction approaches including our proposed INR method, a 1D FFT in readout direction is applied to the 3D k-space volume and image slices are reconstructed individually. All experiments were run on a NVIDIA RTX A2000 12 GB GPU.

\subparagraph{PICS}

For PICS, the TI contrast images are reconstructed directly from the undersampled multi-coil k-space data using the Berkeley Advanced Reconstruction Toolbox \cite{uecker2015bart}. We evaluated $L_2$, wavelet and total variation regularization and selected $L_2$ regularization with $\lambda ~=~ 0.01$ via a parameter sweep.

\subparagraph{Subspace reconstruction}

For subspace reconstruction, the TI contrast image series is constrained to lie in a low-dimensional temporal basis
\(\mathbf{\Phi}\in\mathbb{C}^{N\times K}\) obtained from the signal dictionary. The reconstructed image series
\(\mathbf{d}\in\mathbb{C}^{(V_y\times V_z)\times N}\) is represented as
\begin{equation}
\mathbf{d} ~=~ \boldsymbol{\alpha}\mathbf{\Phi}^{T},
\qquad
\boldsymbol{\alpha}\in\mathbb{C}^{(V_y\times V_z)\times K},
\end{equation}
where \(\boldsymbol{\alpha}\) contains the spatial coefficient images.

The coefficient images are estimated by solving the $L_2$-regularized linear least-squares problem
\begin{equation}
\hat{\boldsymbol{\alpha}}_{\mathrm{L2}}
~=~
\arg\min_{\boldsymbol{\alpha}}
\sum_{c~=~1}^{C}
\left\|
\mathbf{M}\mathbf{F}\mathbf{S}_c
\boldsymbol{\alpha}\mathbf{\Phi}^{T}
-
\mathbf{D}_c
\right\|_2^2
+
\lambda
\left\|
\boldsymbol{\alpha}
\right\|_2^2 .
\end{equation}

solved with conjugate-gradient descent. We selected the $L_2$ regularization parameter as \(\lambda~=~10^{-2}\) with a sweep and evaluated different numbers of subspace bases $K$.

\subparagraph{Iterative Local Low Rank (LLR) reconstruction}

For iterative LLR reconstruction we followed the approach suggested by Allen et al. \cite{allen2026cartesian} minimizing the nuclear norm of $32 \times 32$ neighborhoods around each voxel \cite{Trzasko2011_CLEAR}, using the same regularization parameter and additional denoising through global singular value thresholding.

\subsubsection{Image processing and evaluation metrics}

First, we evaluate the reconstruction methods using fully sampled 1.5 $\mathrm{mm}^3$ data. Next, we assess their performance by retrospectively undersampling this 1.5 $\mathrm{mm}^3$ data. Additional to qualitative assessment, we compute reference-based metrics peak signal to noise ratio (PSNR) and structural similarity index measure (SSIM), relative to reference images reconstructed from the fully sampled data using inverse FFT, for quantitative comparison. Both metrics are calculated with respect to the corresponding reference image, obtained by applying iFFT to the fully sampled data. Before metric calculation, background masking is applied for the phantom measurement, an HD-BET brain mask \cite{Isensee2019} is applied for the volunteer scans, and percentile normalization is applied to both the reference and accelerated images.

Finally, we reconstruct prospectively accelerated 1 $\mathrm{mm}^3$ data using the most promising methods identified from the retrospective evaluation. All images are bias corrected using the N4 bias-field correction algorithm implemented in ANTs \cite{tustison2010n4itk}.

For T1 mapping we use the iterative dictionary described in Section \ref{Iter_dict} with an inversion efficiency of IE ~=~ 1 and IE ~=~ 0.85 for phantom and in vivo measurements, respectively. For visualization of T1 maps we use the lipari colormap \cite{Fuderer2023}.

To assess the T1 fitting accuracy for phantom data we calculate the T1 percentage error as

\begin{equation}
\frac{\mathrm{T1}_{\mathrm{MPnRAGE}} - \mathrm{T1}_{\mathrm{reference}}}
{\mathrm{T1}_{\mathrm{reference}}} \cdot 100 \ [\%]
\end{equation}

with respect to the fully sampled iFFT reference for error maps and with respect to the values from the NIST/ISMRM phantom documentation for the quantitative analysis.

The parameters of our MPnRAGE experiments were determined to fit the timings of a clinically used MPRAGE product sequence. The TI 6 contrast samples the center of k-space, which determines the image contrast, at 984 ms, making it the closest match to the TI ~=~ 1000 ms used in the product MPRAGE sequence. Therefore, TI 6 can be used like a standard T1-weighted image in clinical settings.

\section{Results}\label{sec3}

\subsection{Phantom data}

\subparagraph{Fully sampled data}

We first evaluate the proposed INR and comparison methods (zero-filled iFFT, PICS, subspace, iterative LLR) on fully sampled phantom data. Supporting Information Figure S1 shows high image quality for all methods, with PSNR values above 29 dB and SSIM above 0.92. For subspace reconstruction with $K~=~2$, the contrast at TI 1 differs from the iFFT reference, particularly in the central short-T1 vials.

\begin{figure}[t]
\centerline{\includegraphics[width=\linewidth]{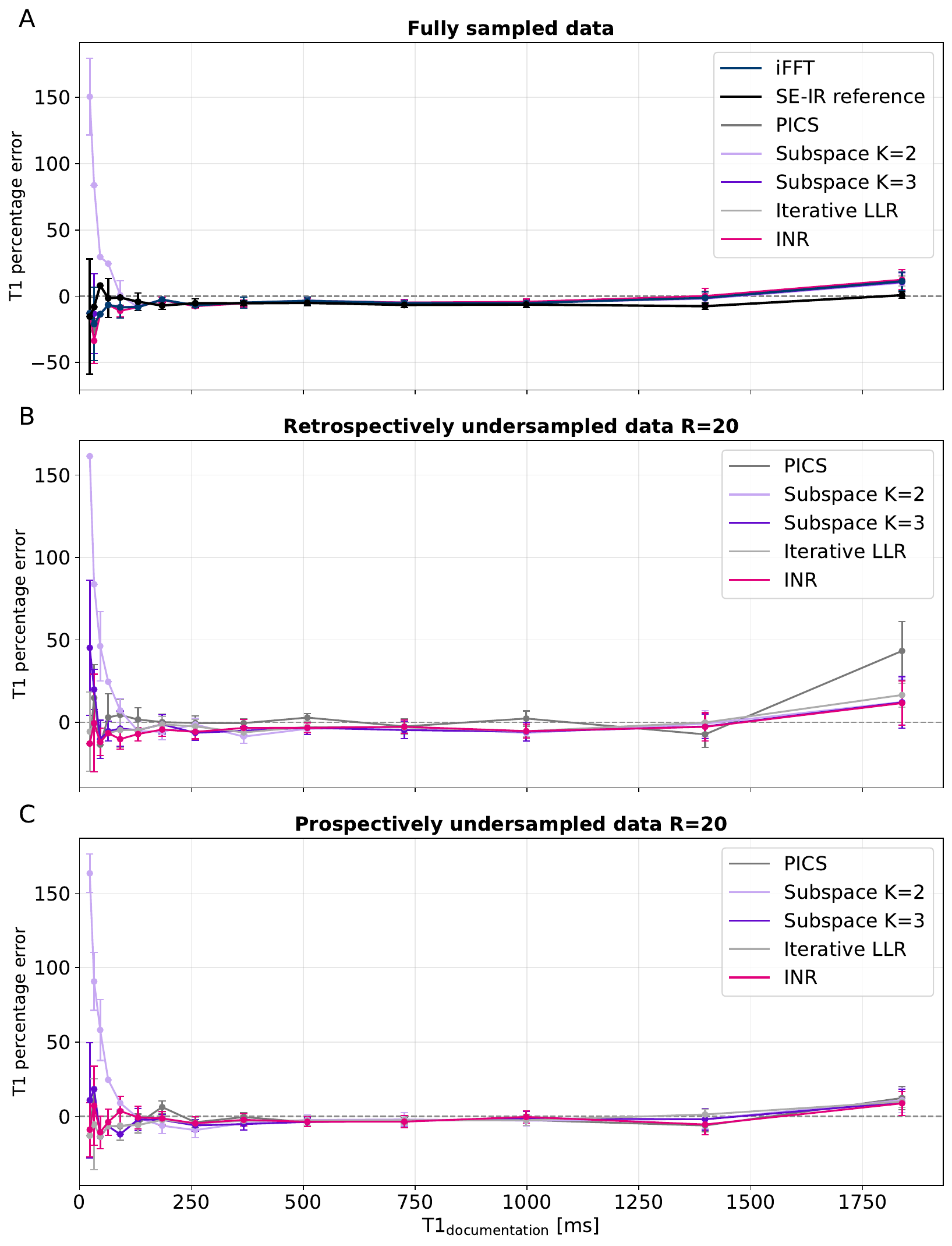}}
\caption{\hspace*{0.5em}
T1 percentage error calculated with respect to the documentation T1 values for 14 T1 vials of NIST/ISMRM phantom data acquired with the proposed MPnRAGE sequence and reconstructed with different reconstruction methods. A) For fully sampled data, subspace reconstruction using K ~=~ 2 bases results in a very high error compared to iFFT, PICS, subspace with K ~=~ 3 and INR reconstruction. All reconstructions show stronger deviations in T1 estimations for very short and long T1. Mid-T1 values are slightly underestimated for all T1 values, which is consistent with an SE-IR scan which serves as a reference.
B) Retrospectively undersampled and C) prospectively accelerated data at R ~=~ 20 showing similar behavior as A.
}\label{fig_T1_fitting}
\end{figure}

Figure~\ref{fig_T1_fitting}A shows the percentage error of T1 estimates for the 14 NIST/ISMRM phantom vials. The shortest and longest T1 values show increased bias and variability, with short-T1 errors up to 150 \% for subspace reconstruction with $K~=~2$ basis functions. The SE-IR reference helps assess deviations. For intermediate T1 values, all methods slightly underestimate the documented values, consistent with this reference.

\begin{figure*}[t]
\centerline{\includegraphics[width=\linewidth]{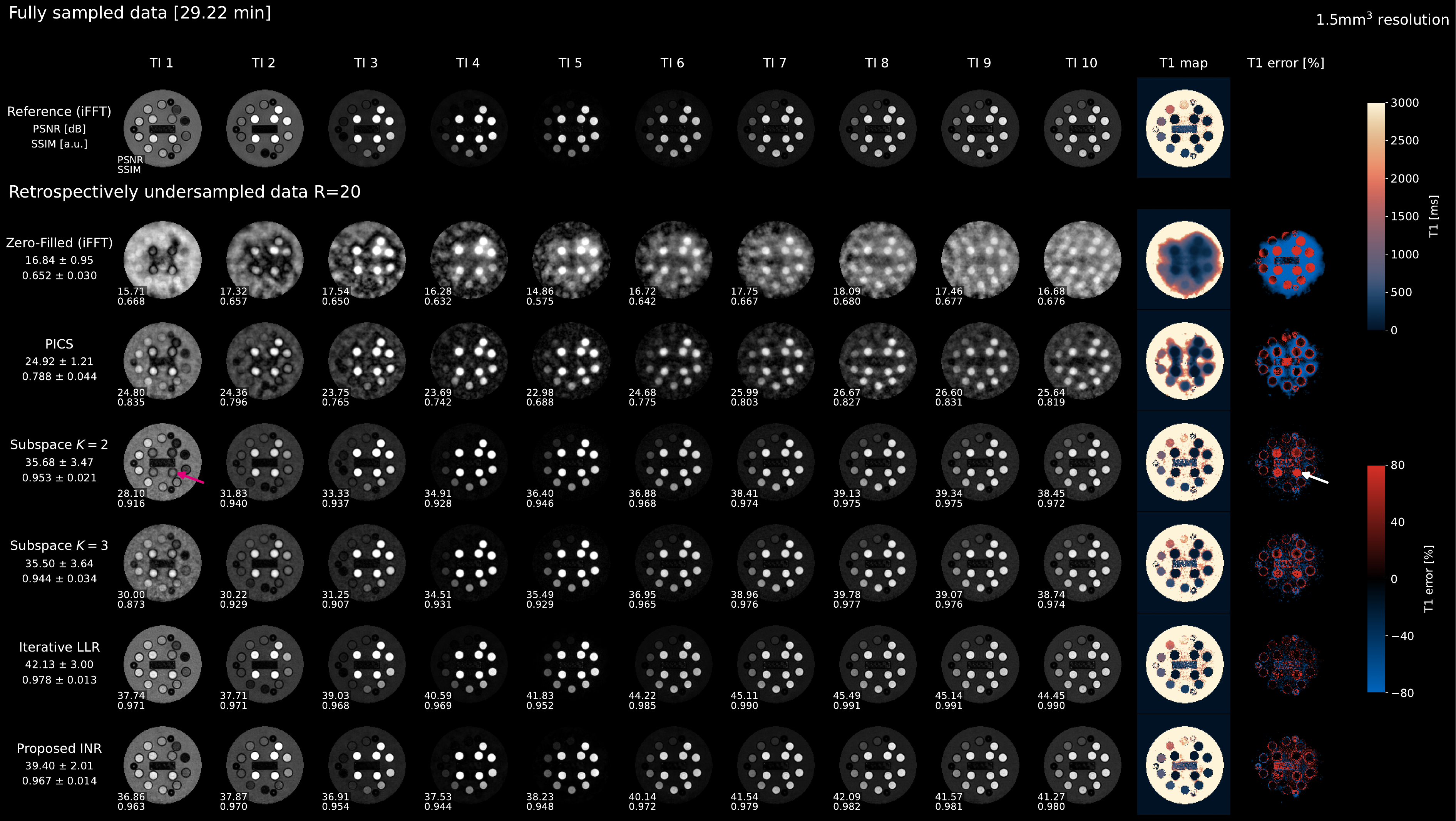}}
\caption{\hspace*{0.5em}
NIST/ISMRM phantom MPnRAGE data and T1 maps at 1.5 $\mathrm{mm}^3$ resolution. Fully sampled data (29.22 minutes scan time) reconstructed with iFFT serves as a reference for calculating PSNR and SSIM metrics, as well as T1 percentage error maps.
Retrospectively undersampled data at R ~=~ 20 show unusable image reconstructions from zero-filled iFFT and PICS. Subspace reconstruction with K ~=~ 2 subspace bases leads to difference in contrast in TI 1 with respect to the reference for short T1 vials also reflected in the T1 error map. With K ~=~ 3 subspace bases TI 1 and TI 2 are blurry. Both iterative LLR and the proposed INR method reconstruct sharp images with contrast similar to the reference and achieve high PSNR and SSIM values. The fully sampled reference exhibits ringing around the T1 vials which is effectively removed by the INR reconstruction as can also be seen in the error map.
}\label{fig_phantom_retro}
\end{figure*}

\subparagraph{Retrospectively undersampled data}

Figure~\ref{fig_phantom_retro} shows reconstructed images and T1 maps for 1.5 $\mathrm{mm}^3$ phantom data retrospectively undersampled with R ~=~ 20. Zero-filled iFFT and PICS show strongly degraded image quality with average PSNR 15.85 and 24.80 dB, and SSIM 0.782 and 0.537, respectively. For subspace reconstruction, $K~=~3$ preserves vial contrast more consistently with the reference, but reduces apparent image quality and introduces edge blurring at TI 1. In contrast, $K~=~2$ improves visual image quality, but again produces a contrast mismatch for short-T1 vials in the phantom center at TI 1, also reflected in the T1 error map. Due to this mismatch, and in accordance with the SVD-based analysis presented in our previous work \cite{Niessen2025_ISMRM}, we use K ~=~ 3 basis functions for all subsequent experiments. Both iterative LLR and INR show sharp images with reference-like contrast and high PSNR and SSIM. The T1 error map shows that the ringing that appears in the fully sampled reference is removed by the INR reconstruction.

The T1 percentage error in Figure~\ref{fig_T1_fitting}B shows trends similar to fully sampled data, with larger method-dependent variation for mid-T1 values. Even with $K~=~3$ subspace bases, the mean percentage error approaches 50 \%. PICS shows stronger overestimation for long T1 values.

\subparagraph{Prospectively undersampled data}

Figure S2 and Figure \ref{fig_T1_fitting}C show similar behavior for prospectively accelerated 1 $\mathrm{mm}^3$ data at R ~=~ 20.

\subsection{In vivo data}

\subsubsection{Acquisition aspects}

\subparagraph{Flip angle schedule}

Figure~\ref{fig_FA_schedule}C demonstrates that our custom FA schedule increases the signal for TI 2 - TI 5, which initially suffered from very low SNR, resulting in a more pronounced WM-GM contrast difference. Tissue-specific segmentation with ANTS \cite{Avants2011} (Figure~\ref{fig_FA_schedule}E) corresponds well with the WM-GM contrast difference calculated from the iterative dictionary (Figure~\ref{fig_FA_schedule}D).

\subparagraph{Complementary Poisson-disk sampling}

\begin{figure*}[t]
\centering
\includegraphics[width=\textwidth]{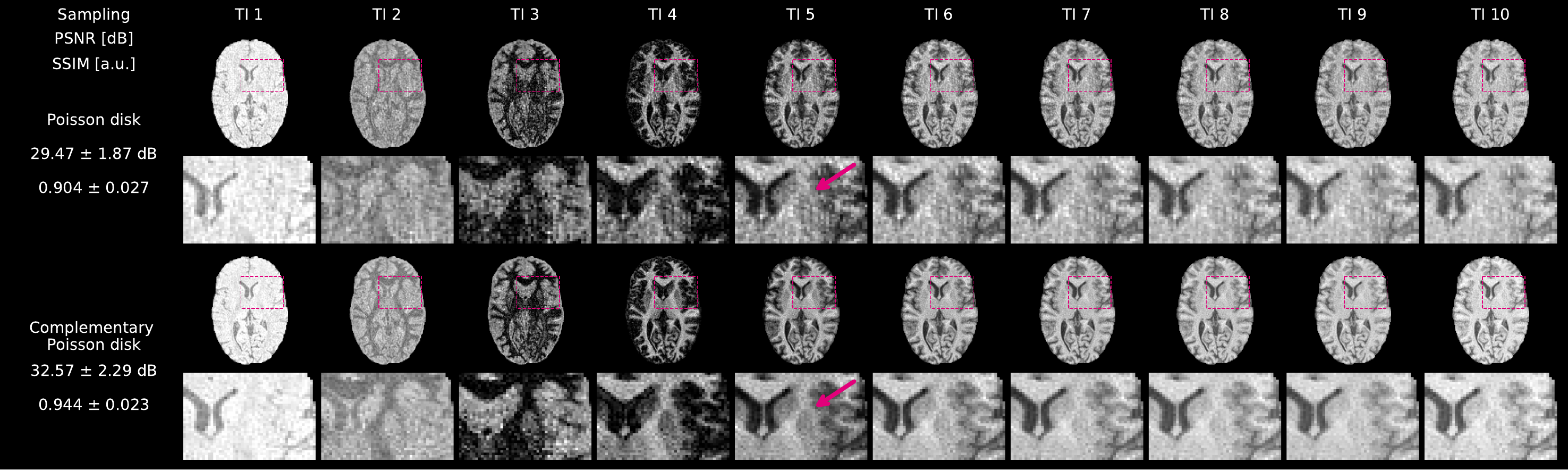}
\caption{\hspace*{0.5em}
In-vivo MPnRAGE data and T1 maps at 1.5 $\mathrm{mm}^3$ resolution, retrospectively undersampled with: Poisson-disk sampling where each TI contrast is sampled with the same sampling mask (top) vs. complementary Poisson-disk undersampling (bottom). Complementary Poisson-disk sampling allows to preserve detailed image structures like the putamen better and achieves higher PSNR and SSIM calculated with respect to a fully sampled iFFT reference than non-complementary Poisson-disk sampling.
}
\label{fig_comp_PD}
\end{figure*}

Figure \ref{fig_comp_PD} shows improved image quality and sharpness using complementary Poisson-disk sampling, with average PSNR 32.57 dB and SSIM 0.944, compared to non-complementary Poisson-disk sampling with PSNR 29.47 and SSIM 0.904.

\subsubsection{Evaluation of reconstruction and T1 mapping}

\subparagraph{Fully sampled data}
In Figure S3, for 1.5 $\mathrm{mm}^3$ fully sampled in-vivo data, all methods perform well, with images and T1 maps closely resembling the reference and achieving PSNR above 37 dB and SSIM above 0.98. PICS ranks highest, followed by subspace, iterative LLR, and INR.

\subparagraph{Retrospectively undersampled data}

\begin{figure*}[t]
\centerline{\includegraphics[width=\linewidth]{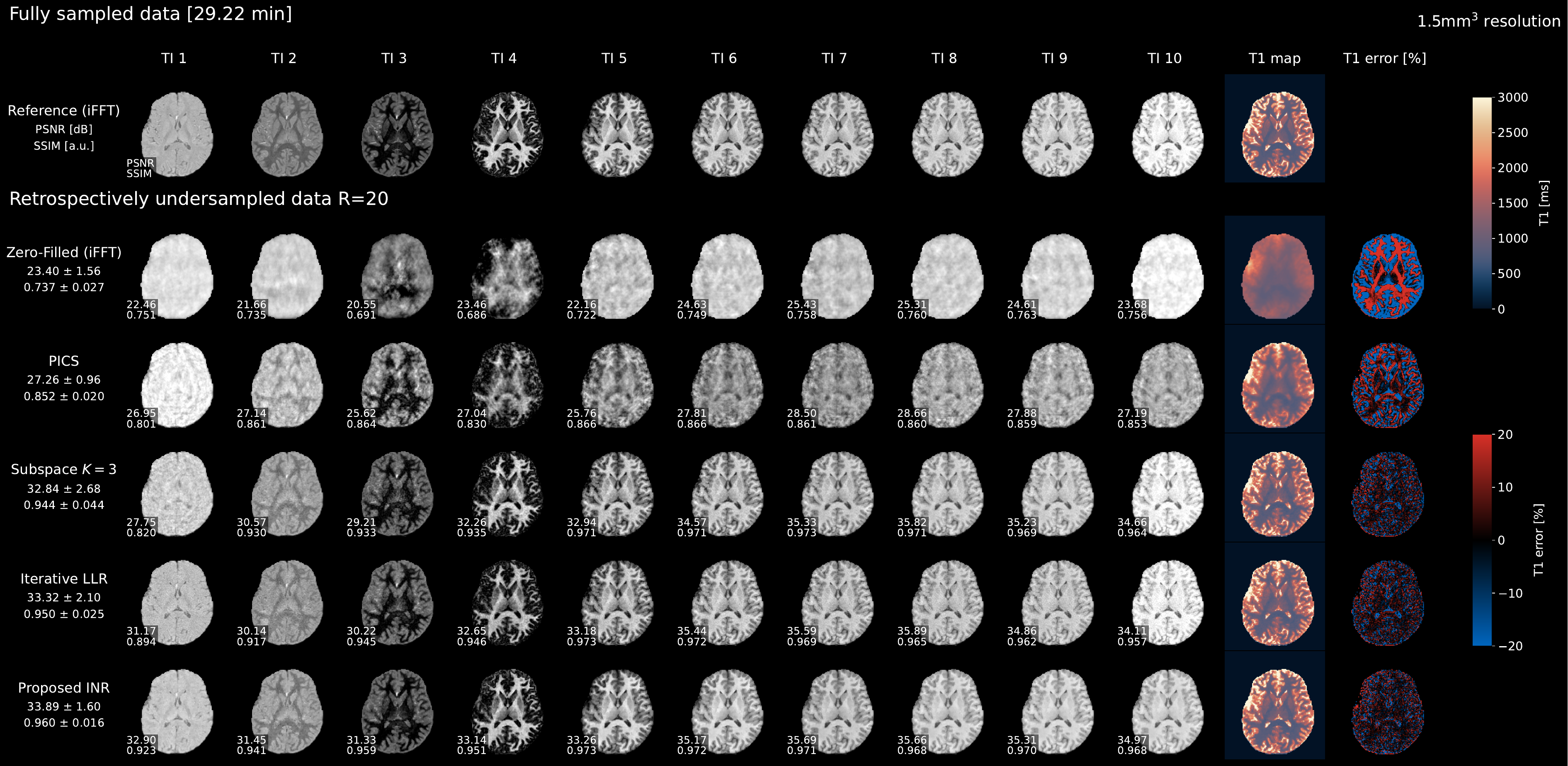}}
\caption{\hspace*{0.5em}In-vivo MPnRAGE data and T1 maps at 1.5 $\mathrm{mm}^3$ resolution. Fully sampled data (29.22 minutes scan time) reconstructed with iFFT serves as a reference for calculating PSNR and SSIM metrics, as well as percentage error maps.
Retrospectively undersampled data at R ~=~ 20 show unusable image reconstructions from zero-filled iFFT and PICS. Subspace reconstruction images TI 1 exhibits artifacts and lower PSNR and SSIM, TI 2 - TI 4 are noisier than INR reconstruction. The proposed INR method and iterative LLR both reconstruct sharp images for all TI with INR achieving the highest PSNR and SSIM values. T1 error maps show the highest error for CSF which has long T1 values.}\label{fig_vivo_retro}
\end{figure*}

The reconstruction results for retrospectively undersampled 1.5 $\mathrm{mm}^3$ data at R ~=~ 20 (Figure~\ref{fig_vivo_retro}) show that simple iFFT fails to reconstruct the underlying tissue information. PICS is also substantially degraded, with reduced metrics (PSNR: 27.26 dB, SSIM: 0.852) relative to fully sampled PICS and strong edge blurring in the T1 map.

Subspace, iterative LLR, and INR reconstruction yield overall good-quality images and T1 maps with similar T1 error maps not showing any strong structural errors except for cerebrospinal fluid (CSF) with long T1 values. However, the subspace-reconstructed image is very blurry for the first TI and noisier especially for TI 2 and TI 3, consistent with lower PSNR and SSIM values for these contrasts (e.g. for TI 1: 27.75 dB/0.820 for subspace vs. 32.93 dB/0.923 for INR).

\begin{figure*}[t]
\centerline{\includegraphics[width=0.6\linewidth]{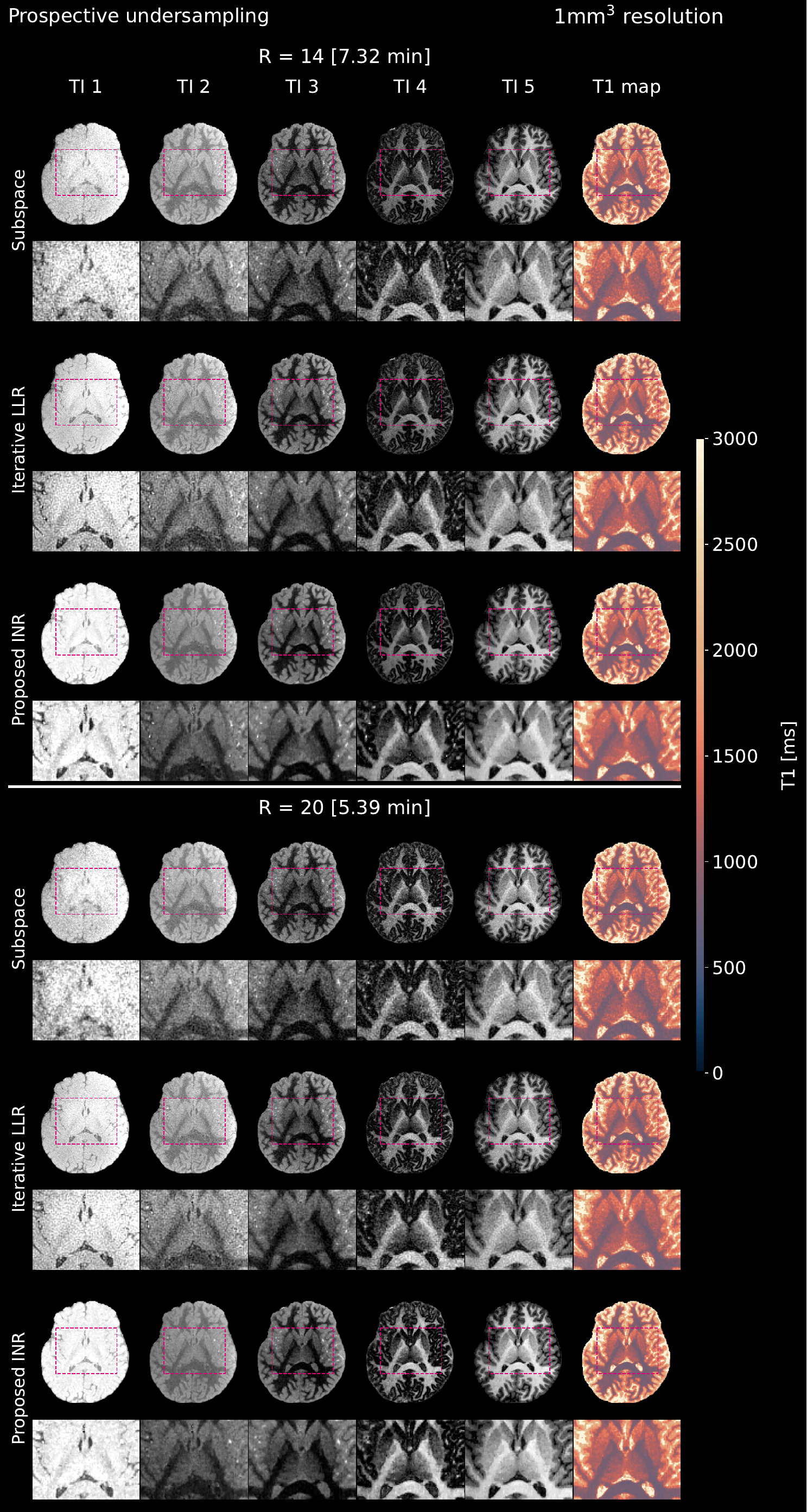}}
\caption{\hspace*{0.5em}
In-vivo MPnRAGE data and T1 maps at 1 $\mathrm{mm}^3$ resolution. Prospectively accelerated data at R ~=~ 14 (7.32 minutes scan time) and R ~=~ 20 (5.39 min) scan time show that the proposed INR framework is robust to increased prospective undersampling while subspace reconstruction exhibits high noise levels at TI 1 and subspace and iterative LLR show increased noise levels at TI 2 for R ~=~ 20 compared to R ~=~ 14. For this high resolution data, no metrics and error maps are available as the fully sampled scan time would have been infeasible.
}\label{fig_vivo_pro}
\end{figure*}

For retrospective undersampling, we additionally evaluate subspace reconstruction, iterative LLR, and INR at increasing undersampling factors. At R ~=~ 20, structures such as the caudate nucleus remain distinguishable from surrounding tissue at TI 2 for subspace reconstruction and iterative LLR (Figure \ref{fig_vivo_pro}), but are largely blurred at TI 2 for R ~=~ 28 and even at TI 3 for R ~=~ 34 (Supportive information Figure S5). INR still represents this structure at TI 2 for R ~=~ 28, with image quality starting to decrease at R ~=~ 34.

\subparagraph{Prospectively accelerated data}

We demonstrate the proposed Cartesian MPnRAGE sequence at the clinically more valuable resolution of 1 $\mathrm{mm}^3$. At R ~=~ 14 (7.32 minutes scan time) and R ~=~ 20 (5.39 minutes), we compare the three most promising techniques from retrospective undersampling: subspace, iterative LLR, and INR. Figure \ref{fig_vivo_pro} shows only the first five TIs because differences mainly occurred for early TI contrasts (Figure \ref{fig_vivo_retro}). All three methods provide good reconstruction results for TI 3 and higher, whereas subspace reconstruction and iterative LLR show increased noise at TI 2 for R ~=~ 20. For TI 1, INR recovers more details, while subspace reconstruction delivers very noisy images. Overall, INR proves robust at increasing prospective accelerations.

The T1 maps of subspace reconstruction and iterative LLR are also slightly noisier than those obtained from INR reconstructed images.

\subsubsection{MPnRAGE derived imaging}

\begin{figure*}[t]
\centerline{\includegraphics[width=\linewidth]{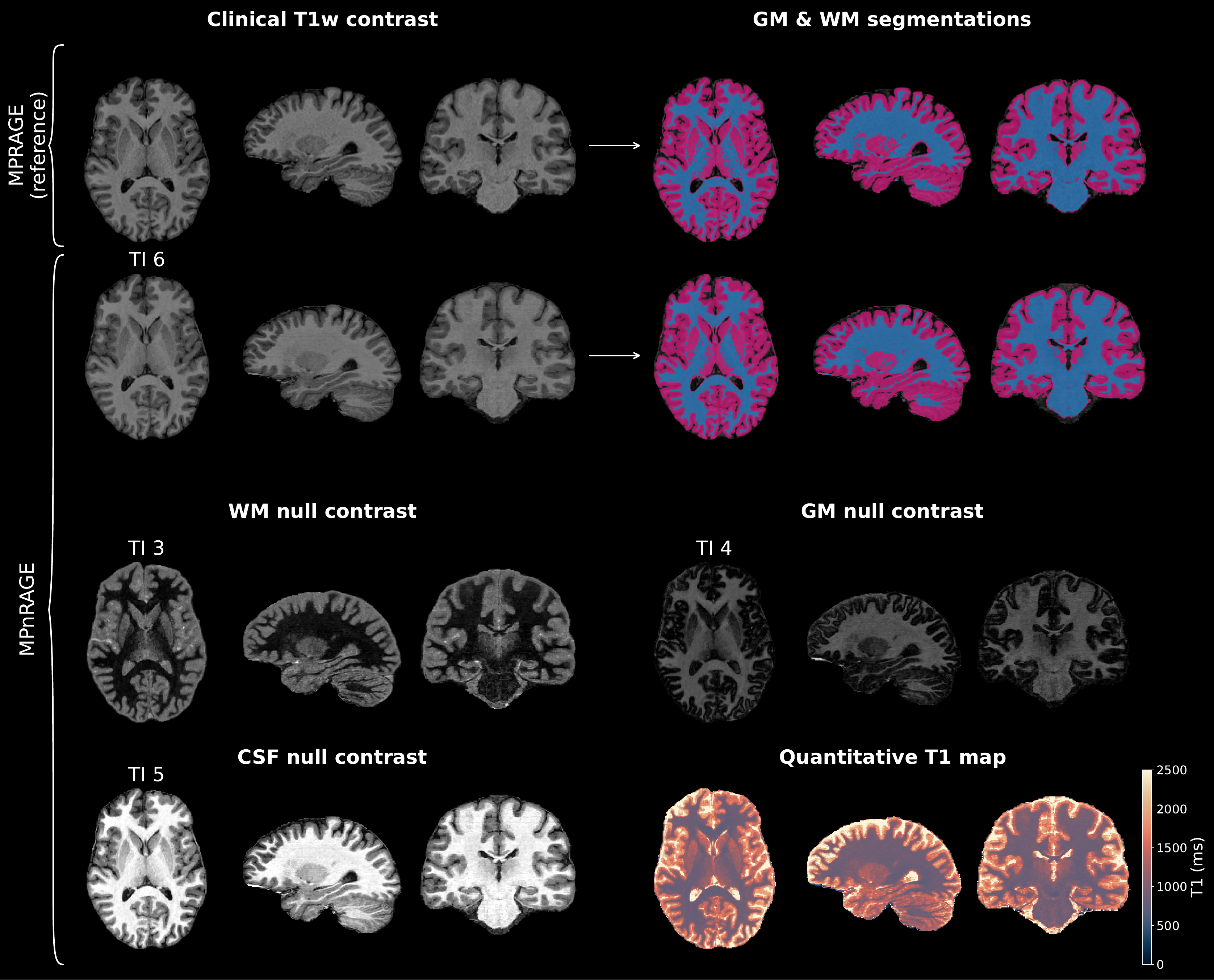}}
\caption{\hspace*{0.5em}
Various contrasts produced by MPnRAGE: MPnRAGE TI 6 ~=~ 984 ms compares well to the product MPRAGE sequence (TI ~=~ 1000 ms) providing a clinical T1-weighted contrast. Gray matter (GM) \& white matter (WM) segmentations obtained from both MPnRAGE TI 6 and MPRAGE show high correspondence. Additionally, MPnRAGE provides tissue-nulled contrasts. At TI 3 WM is null (similar to FLAWS 1), at TI 4 GM is null and at TI 5 cerebrospinal fluid (CSF) is null (similar to FLAWS 2). Finally, quantitative T1 maps can also be obtained from MPnRAGE. All MPnRAGE images are in-vivo 1 $\mathrm{mm}^3$ resolution data acquired at R ~=~ 20 (5.39 minutes scan time) and reconstructed with our proposed INR reconstruction.
}\label{fig_MPRAGE}
\end{figure*}

In Figure \ref{fig_MPRAGE}, we compare the MPRAGE image to TI 6 of the 1$\mathrm{mm}^3$, R ~=~ 20 MPnRAGE. The contrast in both images is very similar and anatomically relevant structures such as the putamen are visualized well. We additionally visualize GM and WM segmentations obtained with ANTS \cite{Avants2011} based on MPRAGE and MPnRAGE TI 6. Furthermore, a WM null contrast (TI 3), GM null contrast (TI 4) and CSF null contrast (TI 5) are shown alongside a quantitative T1 map.

\section{Discussion}\label{sec4}

\subsection{Acquisition aspects}

Our results support the practical viability of the proposed 3D Cartesian MPnRAGE sequence for highly accelerated multi-TI imaging. Compared with radial MPnRAGE acquisitions \cite{Kecskemeti2016_MPnRAGE}, the Cartesian readout enables a 1D FFT along the readout direction and reduces the INR problem to 2D slice-wise reconstructions, avoiding prohibitively large joint volumetric representations \cite{feng2025refine,lao2025summit}.

We leverage previously unused delay periods to acquire scan-matched calibration data. Because these data are acquired after further longitudinal recovery, they provide sensitivity maps that reduce phase errors associated with magnetization regrowth. They also enable background phase correction and provide more dense sampling directly leveraged by the joint INR reconstruction.

Unlike approaches that use an independent center-out trajectory for each TI bin \cite{allen2026cartesian}, our method continuously traverses k-space across successive TI bins with alternating direction (see Figure \ref{fig_sampling}). This minimizes abrupt phase-encoding gradient transitions. In our previous fully sampled center-out Cartesian data \cite{Niessen2025_MICCAI}, cortical ringing artifacts were observed that are not appearing for our proposed sampling strategy. The complementary versus non-complementary Poisson-disk comparison further shows that INR can leverage information across all TI.

We propose a variable flip-angle schedule that increases signal for inherently low-SNR TI contrasts. Although systematic optimization remains necessary, these results indicate that variable flip-angle design can improve SNR and image quality in multi-inversion-recovery imaging.

\subsection{Evaluation of reconstruction and T1 mapping}

We evaluate reconstruction performance and T1 mapping accuracy at high accelerations using fully sampled, retrospectively undersampled, and prospectively accelerated phantom and in-vivo data. Fully sampled experiments show that all methods provide high image quality when sufficient k-space information is available. However, highly accelerated experiments show that methods reconstructing each TI independently, such as zero-filled iFFT and PICS, are not sufficient to obtain usable images, highlighting the importance of jointly exploiting shared information across TI.

The phantom experiments further demonstrate that, for subspace reconstruction, the number of subspace bases defines a tradeoff between denoising and accurate representation of temporal signal evolution, especially for short T1 values where signal recovery changes rapidly across TIs. Using two subspace bases results in visually less noisy images, but causes a pronounced contrast mismatch and overestimation of short T1 values, even for fully sampled data. Using three subspace bases better represents the signal evolution consistent with our previous SVD-based analysis \cite{Niessen2025_ISMRM}, but results in noisier and less sharp images under undersampling (Figure \ref{fig_phantom_retro}). In contrast, INR preserves image details at high acceleration while maintaining more accurate T1 estimates, particularly for very short T1 values.

For in-vivo data, coherent trends between retrospectively undersampled and prospectively accelerated acquisitions indicate that retrospective undersampling provides a useful proxy for prospective performance. The experiments on highly accelerated retrospective undersampling (Figure S6) show that INR is more robust to undersampling and better preserves anatomical structures than iterative LLR and subspace reconstruction, especially for the shorter TIs. This is relevant because short TIs contain important information about rapidly recovering signal components. While subspace reconstruction constrains signal evolution to a dictionary-derived low-rank signal model, INR does not rely on an explicit dictionary or a single-compartment Bloch model during reconstruction. This reduced model dependence may help with partial-volume effects, since single-compartment dictionary models can fail when multiple tissue compartments contribute within a voxel \cite{tang2018multicompartment}, particularly at short TI where rapidly relaxing components such as myelin water can contribute to the signal \cite{liao2023high}.

Self-supervised and scan-specific deep learning-based reconstruction methods, such as the proposed INR, are particularly useful for multi-contrast MRI sequences where large training datasets are not available. The INR was applied with fixed hyperparameters across all subjects, spatial resolutions, and acceleration factors, without case-specific retuning. Figure S7 shows reconstructions for all five volunteers, suggesting robustness across acquisition settings and practical generalizability.

The deep factor model proposed for radial MPnRAGE \cite{Chen2025_RadialMPnRAGE} is another self-supervised approach, however, to the best of our knowledge, no public implementation is currently available and the reported reconstruction time is long, with 4.2 hours for a 9 minute scan (even on an NVIDIA A100 GPU, which is substantially more powerful than the NVIDIA RTX A2000 used in the present study). In contrast, our INR requires approximately 3.45 s per slice, corresponding to about 13.8 minutes for a 1 $\mathrm{mm}^3$ whole-brain undersampled volume (5.39 minute scan), and may be further accelerated through subject-specific pretraining \cite{shen2024nerp} or generalizable priors \cite{li2023efficient}. For comparison, subspace reconstruction took approximately 6.6 minutes and iterative LLR 1.23 hours for the same volume (Table S1).

\subsection{MPnRAGE derived imaging}

The comparison with the clinical product MPRAGE acquisition highlights the versatility of MPnRAGE. From a single acquisition in 5.39 minutes, MPnRAGE provides quantitative T1 maps, several tissue-nulled contrasts, and a conventional T1-weighted image contrast (see Figure \ref{fig_MPRAGE}) which itself already takes 4.26 minutes to acquire. MPnRAGE yields suppression of WM at TI 3, GM at TI 4, and CSF at TI5. The latter two correspond to the FLAWS images that can be obtained from MP2RAGE.

Despite minor appearance differences from the vendor product reconstruction, the MPnRAGE-derived T1-weighted image closely resembles the clinically established MPRAGE contrast. If an exact match to MPRAGE or another timing is desired, the target contrast could be synthesized from estimated T1 maps, although partial-volume effects may not be fully preserved. Alternatively, image-domain synthesis using a linear combination of acquired TI images could more directly approximate the desired timing.

The T1 fitting results demonstrate that the proposed Cartesian MPnRAGE sequence together with INR reconstruction enables quantitative T1 estimation. In phantom experiments, longer T1 values were slightly overestimated, most likely due to incomplete longitudinal recovery before subsequent inversion pulses. A similar tendency was observed in vivo, particularly in GM. Nevertheless, the reconstructed T1 maps show consistent tissue contrast and support MPnRAGE as a combined quantitative and qualitative imaging approach.

\subsection{Limitations \& Future directions}

Reconstructing whole high-resolution, multi-contrast, multi-coil MRI volumes with an INR remains challenging in random access memory (RAM) usage and computation time. Therefore, the INR reconstructs each slice individually, enabled by the Cartesian undersampling pattern and a 1D Fourier transform along the readout direction before INR-based reconstruction. However, slice-wise processing may introduce subtle slice-to-slice artifacts, as indicated in TI 1 in Figure S4. These artifacts are expected to be mitigated by extending the method to a 3D INR reconstruction, which represents a natural direction for future work.

Another limitation is reconstruction time. Even though INR is comparably fast for a scan-specific, self-supervised method, 13.8 minutes for a whole undersampled R ~=~ 20 volume with 8 to 12 coils is still too long for clinical applications. Reconstructions need to be near-instantaneous to assess image quality and potentially rescan after movement or other disturbances. Future work will therefore focus on further acceleration by pretraining the INR on composite information from all contrasts or improving network initialization.

This study was performed in healthy volunteers only. The accelerated MPnRAGE sequence is currently being used in a study protocol that will enable evaluation in patient data. Clinically, multiple tissue-nulled contrasts may improve segmentation of smaller brain structures such as thalamic nuclei \cite{Datta2021_segmentation} or brainstem nuclei \cite{Mueller2020_MP2RAGEBrainstem}, lesion characterization, and tissue-specific assessment \cite{Gkotsoulias2026_ISMRM}.

The proposed reconstruction framework may also extend to quantitative multi-contrast acquisitions such as 3D-QALAS \cite{Kvernby2014_3DQALAS}, which provides T1, T2, and proton-density information. Such multi-parametric acquisitions could enable additional quantitative modeling and potentially support INR-enabled multi-compartment fitting.

The INR reconstruction code is publicly available [github link will be added upon publication] and the MPnRAGE sequence can be shared for collaboration through the GE HealthCare portal (https://weconnect.gehealthcare.com).

\section{Conclusion}
In conclusion, we developed a Cartesian 3D MPnRAGE acquisition with a tailored complementary variable-density Poisson-disk sampling strategy that keeps consecutively acquired TI samples spatially close in k-space while minimizing abrupt changes in phase-encoding gradient amplitude. We leverage the usually unused delay time where the contrast changes are minimal for acquiring calibration data that contributes to the joint reconstruction with even more data and can be used e.g. for estimating sensitivity maps.
The complementary undersampled data are reconstructed jointly using an implicit neural representation (INR), which outperforms state-of-the-art reconstruction methods and provides high-quality images even at high acceleration factors of R ~=~ 20, corresponding to a scan time of 5.39 min. The resulting images exhibit various tissue-nulled contrasts and can be used for quantitative T1 fitting, as well as for generating a T1-weighted image that closely matches the clinically used MPRAGE.

\section*{Acknowledgments}
This work is supported by the DAAD programme Konrad Zuse Schools of Excellence in Artificial Intelligence and the Munich Center for Machine Learning , both sponsored by the Federal Ministry of Research, Technology and Space. The authors thank all PREDICTOM participants and clinical sites for their contribution. This project is supported by the Innovative Health Initiative Joint Undertaking (IHI JU) under Grant Agreement No 101132356. The JU receives support from the European Union’s Horizon Europe research and innovation programme and  Helse Stavanger HF, King’s College London, Foundation Lygature, Ethnniko Kentro Erevnas kai Technolgikis Anaptyxis - Centre for Research and Technology Hellas, Fraunhofer-Gesellschaft zur Förderung der angewandten Forschung e.V., JOANNEUM RESEARCH Forschungsgesellschaft mbH., Qairnel SAS, Ludwig-Maximilians-Universität München, LMU Klinikum München, National Institute for Health and Care Excellence, Alzheimer Europe, Pharmacoidea Fejlesztő és Szolgáltató KFT, Novo Nordisk A/S, GE HealthCare, Siemens Healthineers, The University of Exeter, Icometrix nv, Universitätsklinikum Erlangen, Vrije Universiteit Brussel, Fundacion Para la Investigacion del Hospital Universitario La Fe de la Comunidad Valenciana, ALZpath Inc, GN Hearing AS, Muhdo Health Ltd, Université de Genève, BrainCheck Inc, Neuroelectrics SL, and Starlab Barcelona SL.

The UK participants are supported by UKRI Grant No 10083467 (National Institute for Health and Care Excellence), Grant No10083181 (King’s College London), and Grant No 10091560 (University of Exeter). In Switzerland, the University of Geneva is funded for PREDICTOM by the Swiss State Secretariat for Education, Research and Innovation (SERI – Ref – 1131 52304). Views and opinions expressed are however those of the author(s) only and do not necessarily reflect those of the aforementioned parties. Neither of the aforementioned parties can be held responsible for them.
This work was also supported by the European Research Council (ERC) under Grant Agreement No. 884622.

\bibliography{MRM-AMA}%

@article{mugler1990mprage,
  author  = {Mugler, III, John P. and Brookeman, James R.},
  title   = {Three-dimensional magnetization-prepared rapid gradient-echo imaging ({3D MP RAGE})},
  journal = {Magnetic Resonance in Medicine},
  year    = {1990},
  volume  = {15},
  number  = {1},
  pages   = {152--157},
  doi     = {10.1002/mrm.1910150117}
}

@article{Marques2010_MP2RAGE,
  author = {Marques, José P and Kober, Tobias and Krueger, Gunnar and van der Zwaag, Wietske and Van de Moortele, Pierre-François and Gruetter, Rolf},
  title = {MP2RAGE, a self bias-field corrected sequence for improved segmentation and T1-mapping at high field},
  year = {2010},
  journal = {NeuroImage},
  volume = {49},
  number = {2},
  pages = {1271--1281}
}

@article{Kecskemeti2016_MPnRAGE,
  author = {Kecskemeti, S and Samsonov, A and Hurley, SA and Dean, DC and Field, A and Alexander, AL},
  title = {MPnRAGE: A technique to simultaneously acquire hundreds of differently contrasted MPRAGE images with applications to quantitative T1 mapping},
  year = {2016},
  journal = {Magn Reson Med},
  volume = {75},
  number = {3},
  pages = {1040--1053}
}

@misc{Kecskemeti2015_MS_MPnRAGE,
 author = {Kecskemeti, Steven R. and Alexander, Andrew L. and Field, Aaron S.},
 title = {An 8 Month Longitudinal Study of T1 Measures in Multiple Sclerosis Patients Using 3D MPnRAGE},
 year = {In Proceedings of the Annual Meeting of ISMRM, Toronto, 2015. Abstract 1407.},
}

@article{Hammernik2023,
  author = {Hammernik, Kerstin and Küstner, Thomas and Yaman, Burhaneddin {et al.}},
  title = {Physics-Driven Deep Learning for Computational Magnetic Resonance Imaging: Combining physics and machine learning for improved medical imaging},
  year = {2023},
  journal = {IEEE Signal Process Mag},
  volume = {40},
  number = {1},
  pages = {98--114}
}

@article{Heckel2021,
  author = {Heckel, Reinhard and Soltanolkotabi, Mahdi},
  title = {Unrolled optimization for MRI reconstruction},
  year = {2021},
  journal = {IEEE Signal Process Mag},
  volume = {38},
  number = {1},
  pages = {75--102}
}

@article{Chen2025_RadialMPnRAGE,
  author = {Chen, Yan and Kecskemeti, Steven R. and Holmes, James H. {et al.}},
  title = {Accelerating 3D radial MPnRAGE using a self-supervised deep factor model},
  year = {2025},
  journal = {Magn Reson Med},
  volume = {94},
  number = {3},
  pages = {1191--1201}
}

@article{feng2025refine,
  author = {Feng, Ruimin and Jang, Albert and He, Xingxin and Liu, Fang},
  title = {Accelerating Multiparametric Quantitative MRI Using Self-Supervised Scan-Specific Implicit Neural Representation With Model Reinforcement},
  journal = {Magnetic Resonance in Medicine},
  year = {2025},
  volume = {early access},
  number = {early access},
  doi = {10.1002/mrm.70227}
}

@article{zhang2026lorein,
  author = {Zhang, Haonan and Lao, Guoyan and Zhang, Yuyao and Wei, Hongjiang},
  title = {Unsupervised Highly Accelerated 3D Multi-Parametric MRI Reconstruction via Low-Rank Integrated Implicit Neural Representation},
  journal = {Pattern Recognition},
  volume = {179},
  pages = {113859},
  year = {2026},
  doi = {10.1016/j.patcog.2026.113859}
}

@article{lao2025summit,
  author = {Lao, Guoyan and Feng, Ruimin and Qi, Haikun {et al.}},
  title = {Coordinate-based Neural Representation Enabling Zero-Shot Learning for Fast 3D Multiparametric Quantitative MRI},
  journal = {Medical Image Analysis},
  volume = {102},
  pages = {103530},
  year = {2025},
  doi = {10.1016/j.media.2025.103530}
}

@misc{Niessen2025_MICCAI,
 author = {Niessen, Natascha and Pirkl, Carolin M. and Solana, Ana Beatriz {et al.}},
 title = {INR Meets Multi-contrast MRI Reconstruction},
 year = {In Reconstruction and Imaging Motion Estimation, and Graphs in Biomedical Image Analysis: First International Workshop, RIME 2025, and 7th International Workshop, GRAIL 2025, Berlin, Heidelberg, 2025, pp. 23--33.},
}

@article{allen2026cartesian,
  author = {Allen, Carly A. and Johnson, Kevin M. and Gillick, Bernadette T. and Schimek, Maren E. and Alexander, Andrew L. and Kecskemeti, Steven R.},
  title = {Cartesian {MPnRAGE} for Efficient Simultaneous Multi-Contrast and Quantitative Relaxometry Imaging},
  journal = {Magnetic Resonance in Medicine},
  year = {2026},
  article-number = {mrm.70486},
  doi = {10.1002/mrm.70486}
}

@article{Levine2017_PoissonDisc,
  author = {Levine, Evan G and Daniel, Bruce L and Vasanawala, Shreyas S and Hargreaves, Brian A and Saranathan, Manojkumar},
  title = {3D Cartesian MRI with compressed sensing and variable view sharing using complementary Poisson-disc sampling},
  year = {2017},
  journal = {Magn Reson Med},
  volume = {77},
  number = {5},
  pages = {1774--1785},
  doi = {10.1002/mrm.26254}
}

@article{spieker2024deep,
  title={Deep Learning for Retrospective Motion Correction in {MRI}: A Comprehensive Review},
  author={Spieker, Veronika and Eichhorn, Hannah and Hammernik, Kerstin and Rueckert, Daniel and Preibisch, Christine and Karampinos, Dimitrios C. and Schnabel, Julia A.},
  journal={IEEE Transactions on Medical Imaging},
  volume={43},
  number={2},
  pages={846--859},
  year={2024},
  doi={10.1109/TMI.2023.3323215},
}

@misc{Niessen2026_ISMRM,
 author = {Niessen, Natascha and Solana, Ana Beatriz and Sprenger, Tim {et al.}},
 title = {Highly Accelerated 3D MPnRAGE Meets Implicit Neural Representation (INR) Reconstruction},
 year = {In Proceedings of the Annual Meeting of ISMRM, 2026.},
}

@misc{Niessen2025_ISMRM,
 author = {Niessen, Natascha and Sprenger, Tim and Pirkl, Carolin M. {et al.}},
 title = {Probing the Sparsity of the MPnRAGE Sequence Through Subspace Compression},
 year = {In Proceedings of the Annual Meeting of ISMRM, Honolulu, 2025.},
}

@article{ZavalaBojorquez2017,
  author = {Zavala Bojorquez, Jorge and Bricq, Stéphanie and Acquitter, Clément and Brunotte, François and Walker, Paul M. and Lalande, Alain},
  title = {What Are Normal Relaxation Times of Tissues at 3 T?},
  year = {2017},
  journal = {Magnetic Resonance Imaging},
  volume = {35},
  pages = {69--80}
}

@article{Mueller2022,
  author = {Müller, Thomas and Evans, Alex and Schied, Christoph and Keller, Alexander},
  title = {Instant Neural Graphics Primitives with a Multiresolution Hash Encoding},
  year = {2022},
  journal = {ACM Transactions on Graphics},
  volume = {41},
  number = {4},
  pages = {102:1--102:15}
}

@article{Stupic2021_NISTPhantom,
  author = {Stupic, Karl F. and Ainslie, Maureen and Boss, Michael A. {et al.}},
  title = {A standard system phantom for magnetic resonance imaging},
  journal = {Magnetic Resonance in Medicine},
  year = {2021},
  volume = {86},
  number = {2},
  pages = {1194--1211},
  doi = {10.1002/mrm.28779}
}

@misc{uecker2015bart,
 author = {Uecker, Martin and Ong, Frank and Tamir, Jonathan I. {et al.}},
 title = {Berkeley Advanced Reconstruction Toolbox},
 year = {In Proceedings of the Annual Meeting of ISMRM, Toronto, 2015. Abstract 2486.},
}

@misc{Trzasko2011_CLEAR,
 author = {Trzasko, J D and Manduca, A},
 title = {Calibrationless Parallel MRI Using CLEAR},
 year = {In Proceedings of the 2011 Conference Record of the Forty Fifth Asilomar Conference on Signals, Systems and Computers (ASILOMAR), 2011, pp. 75--79. IEEE.},
}

@article{Isensee2019,
  author = {Isensee, Fabian and Schell, Marianne and Pflueger, Irada {et al.}},
  title = {Automated Brain Extraction of Multisequence MRI Using Artificial Neural Networks},
  year = {2019},
  journal = {Human Brain Mapping},
  volume = {40},
  number = {17},
  pages = {4952--4964},
  doi = {10.1002/hbm.24750}
}

@article{tustison2010n4itk,
  author = {Tustison, Nicholas J. and Avants, Brian B. and Cook, Philip A. {et al.}},
  title = {N4ITK: Improved N3 Bias Correction},
  journal = {IEEE Transactions on Medical Imaging},
  volume = {29},
  number = {6},
  pages = {1310--1320},
  year = {2010}
}

@misc{Fuderer2023,
  author = {Fuderer, Miha},
  title = {mfuderer/colorResources: v1.1 Initial (re-try)},
  year = {2023},
  publisher = {Zenodo},
  version = {1.1},
  doi = {10.5281/zenodo.8268885}
}

@article{Avants2011,
  author = {Avants, Brian B. and Tustison, Nicholas J. and Wu, Jue and Cook, Philip A. and Gee, James C.},
  title = {An Open Source Multivariate Framework for N-Tissue Segmentation with Evaluation on Public Data},
  journal = {Neuroinformatics},
  year = {2011},
  volume = {9},
  number = {4},
  pages = {381--400},
  doi = {10.1007/s12021-011-9109-y}
}

@article{shen2024nerp,
  author = {Shen, Liyue and Pauly, John and Xing, Lei},
  title = {{NeRP}: Implicit Neural Representation Learning With Prior Embedding for Sparsely Sampled Image Reconstruction},
  journal = {IEEE Transactions on Neural Networks and Learning Systems},
  volume = {35},
  number = {1},
  pages = {770--782},
  year = {2024}
}

@article{li2023efficient,
  author = {Li, Hao and Zhou, Yusheng and Liu, Jianan {et al.}},
  title = {Efficient MRI Parallel Imaging Reconstruction by K-Space Rendering via Generalized Implicit Neural Representation},
  journal = {arXiv preprint arXiv:2309.06067},
  year = {2023}
}

@article{tang2018multicompartment,
  author = {Tang, Sunli and Fernandez-Granda, Carlos and Lannuzel, Sylvain {et al.}},
  title = {Multicompartment Magnetic Resonance Fingerprinting},
  journal = {Inverse Problems},
  volume = {34},
  number = {9},
  pages = {094005},
  year = {2018},
  doi = {10.1088/1361-6420/aad1c3}
}

@article{liao2023high,
  author = {Liao, Congyu and Cao, Xiaozhi and Iyer, Siddharth Srinivasan {et al.}},
  title = {High-Resolution Myelin-Water Fraction and Quantitative Relaxation Mapping Using 3D {ViSTa-MR} Fingerprinting},
  journal = {Magnetic Resonance in Medicine},
  year = {2023},
  doi = {10.1002/mrm.29990}
}

@article{Tanner2012_FLAWSMP2RAGE,
 author = {Tanner, Mark and Gambarota, Giulio and Kober, Tobias and Krueger, Gunnar and Erritzoe, David and Marques, Jose P and Newbould, Rexford},
 title = {Fluid and white matter suppression with the MP2RAGE sequence},
 journal = {Journal of Magnetic Resonance Imaging},
 year = {2012},
 volume = {35},
 number = {5},
 pages = {1063--1070},
 doi = {10.1002/jmri.23532},
}

@article{Datta2021_segmentation,
 author = {Datta, R and Bacchus, M K and Kumar, D {et al.}},
 title = {Fast automatic segmentation of thalamic nuclei from MP2RAGE acquisition at 7 Tesla},
 journal = {Magnetic Resonance in Medicine},
 year = {2021},
 volume = {85},
 number = {5},
 pages = {2781--2790},
 doi = {10.1002/mrm.28608},
 pmid = {33270943},
}

@article{Mueller2020_MP2RAGEBrainstem,
 author = {Mueller, S G},
 title = {Mapping internal brainstem structures using MP2RAGE derived T1 weighted and T1 relaxation images at 3 and 7 T},
 journal = {Human Brain Mapping},
 year = {2020},
 volume = {41},
 number = {8},
 pages = {2173--2186},
 doi = {10.1002/hbm.24938},
 pmid = {31971322},
 pmcid = {PMC7198362},
}

@misc{Gkotsoulias2026_ISMRM,
 author = {Gkotsoulias, Dimitrios and Schönenberger, Lukas and Leupold, Jochen {et al.}},
 title = {Multiple Inversion Recovery GRE as a Novel Imaging Marker for Myelin State and Axonal Damage in MS Lesions},
 year = {In Proceedings of the Annual Meeting of ISMRM, 2026. Abstract 404-03-005.},
}

@article{Kvernby2014_3DQALAS,
 author = {Kvernby, S and Warntjes, M J and Haraldsson, H and Carlhäll, C J and Engvall, J and Ebbers, T},
 title = {Simultaneous three-dimensional myocardial T1 and T2 mapping in one breath hold with 3D-QALAS},
 journal = {Journal of Cardiovascular Magnetic Resonance},
 year = {2014},
 volume = {16},
 number = {1},
 pages = {102},
 doi = {10.1186/s12968-014-0102-0},
 pmid = {25526880},
 pmcid = {PMC4272556},
}

\section*{Supporting information}
The following supporting information is available as part of the online article:

\setcounter{figure}{0} 
\setcounter{table}{0} 

\renewcommand{\thefigure}{S\arabic{figure}}

%
\begin{figure*}
\centerline{\includegraphics[width=\linewidth]{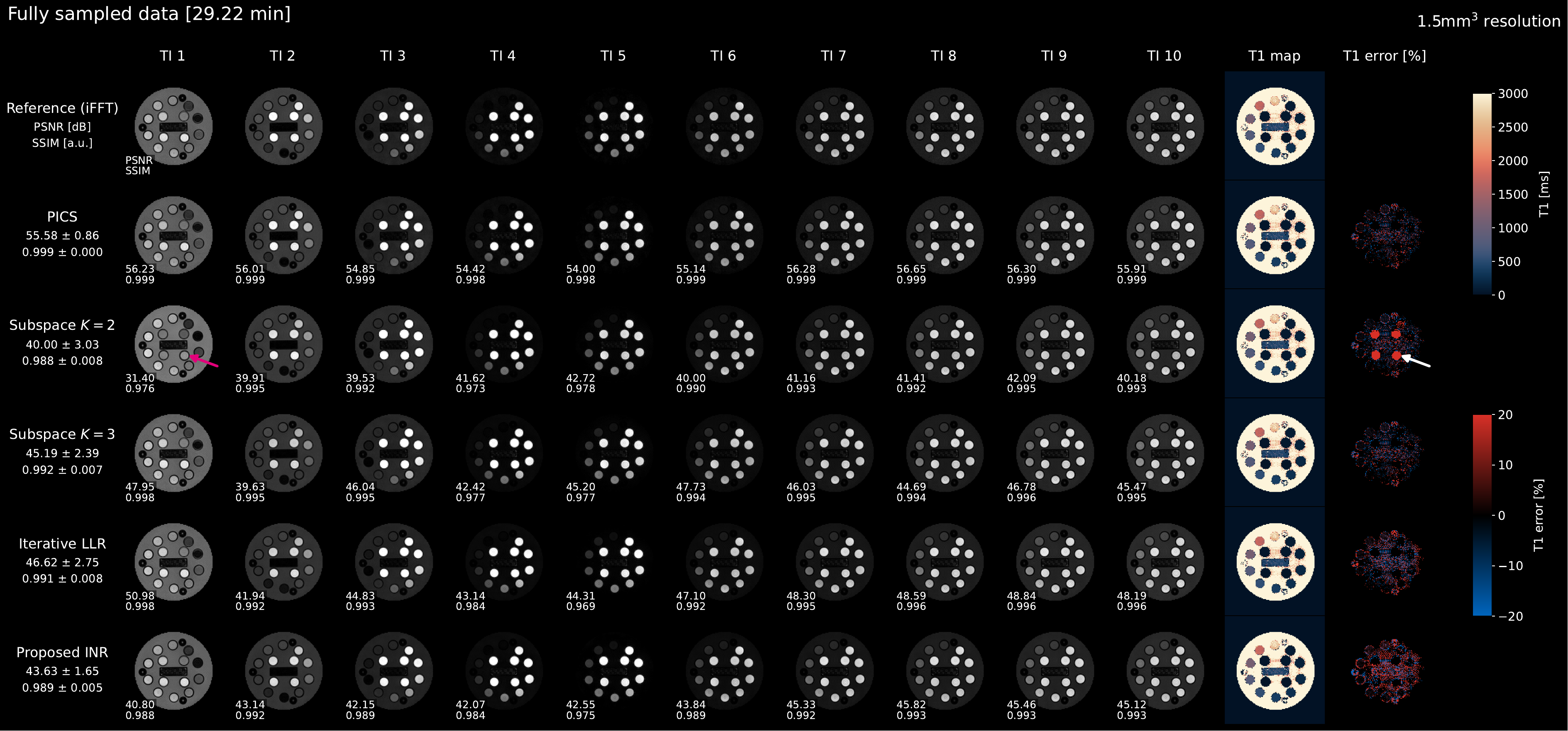}}
\caption{\hspace*{0.5em}{NIST/ISMRM phantom MPnRAGE data and T1 maps at 1.5 $\mathrm{mm}^3$ resolution. Fully sampled data (29.22 min scan time) is reconstructed with iFFT, PICS, Subspace reconstruction using K~=~2 and K~=~3 bases and our proposed INR reconstruction. All reconstructions deliver sharp images with high PSNR and SSIM metrics calculated with respect to the iFFT reference. For subspace with K~=~2, the TI 1 contrast and quantitative fitting of the short T1 vials at the center differs from the reference (marked by the arrows in the figure).
}}\label{fig:Supp_phantom_full}
\end{figure*}
%

%
\begin{figure*}
\centerline{\includegraphics[width=\linewidth]{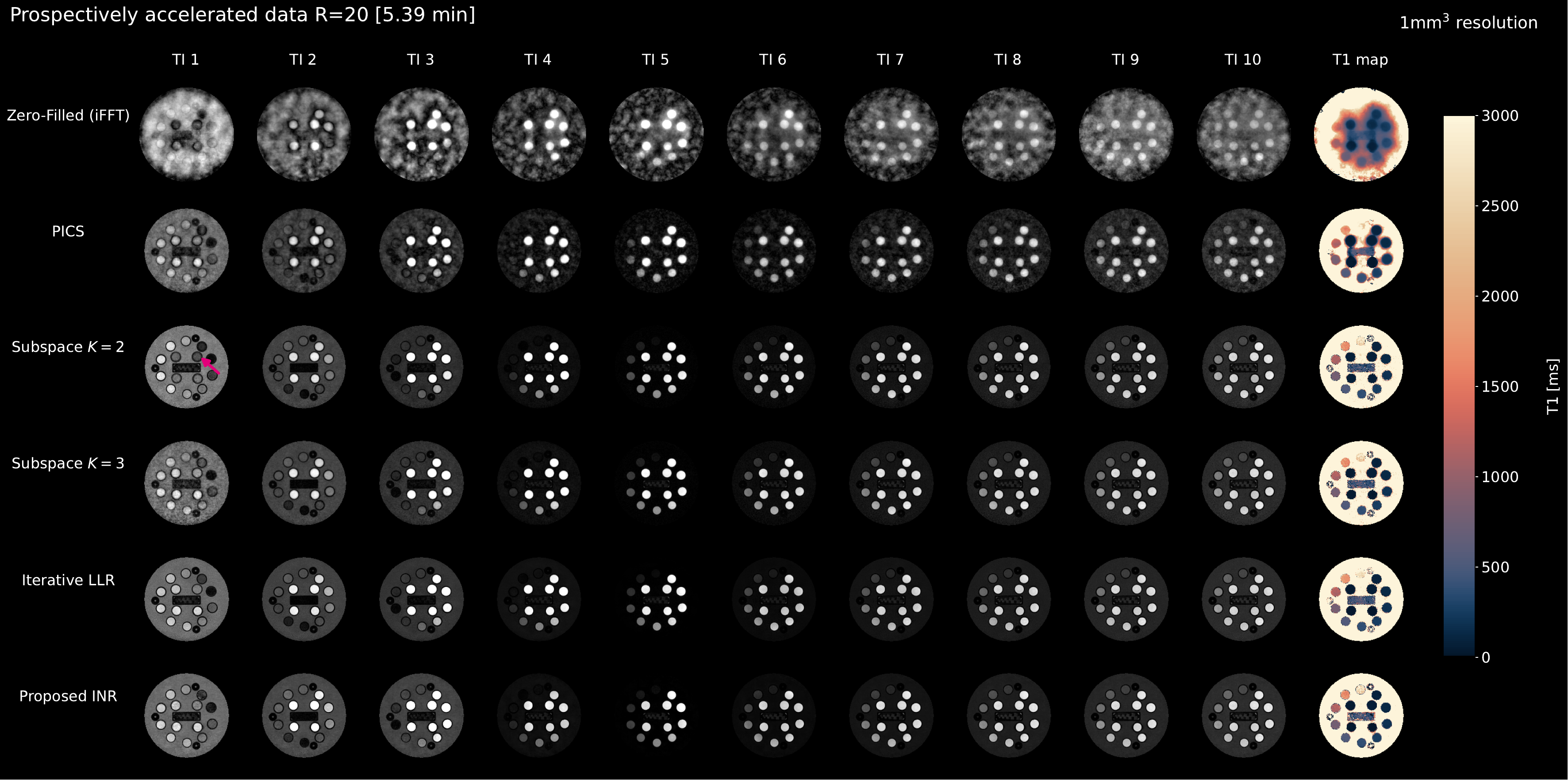}}
\caption{\hspace*{0.5em}{
NIST/ISMRM phantom MPnRAGE data and T1 maps at 1 $\mathrm{mm}^3$ resolution. Prospectively accelerated data (5.39 min) reconstructed with zero-filled iFFT and PICS result in unusable and blurry images, respectively. For subspace with K~=~2, the contrast of the short T1 vials at the center differs from the reference. with K~=~3 subspace bases TI 1 and TI 2 are blurry. Iterative LLR and the proposed INR method reconstruct sharp images with contrast similar to the reference.
}}\label{fig:Supp_phantom_pro}
\end{figure*}
%

%
\begin{figure*}
\centerline{\includegraphics[width=\linewidth]{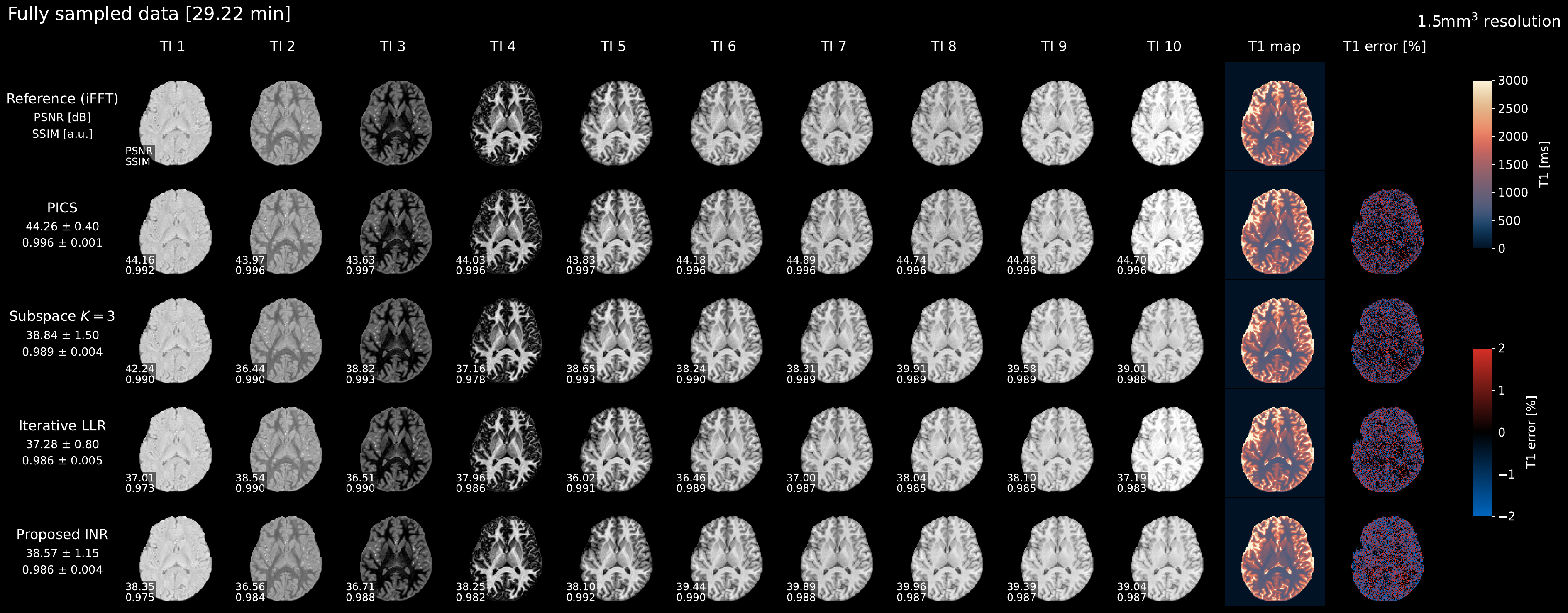}}
\caption{\hspace*{0.5em}{
In-vivo MPnRAGE data and T1 maps at 1.5 $\mathrm{mm}^3$ resolution. Fully sampled data (29.22 min scan time) is reconstructed with iFFT, PICS, subspace, iterative LLR and our proposed INR reconstruction. All reconstructions deliver sharp images with high PSNR and SSIM metrics calculated with respect to the iFFT reference.
}}\label{fig:volunteer_fullysampled}
\end{figure*}
%

%
\begin{figure*}
\centerline{\includegraphics[width=\linewidth]{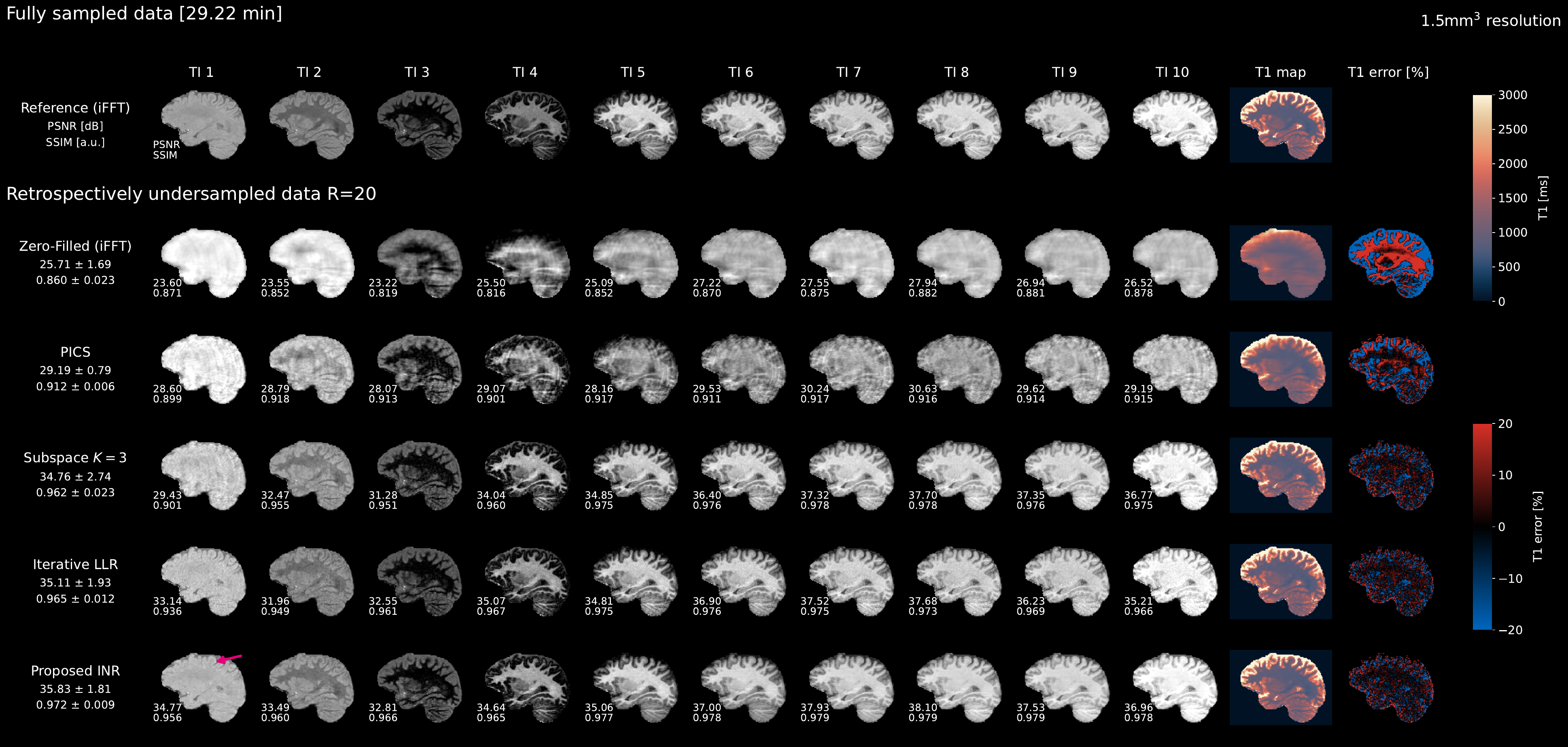}}
\caption{\hspace*{0.5em}{
Retrospective undersampling demonstrated for sagittal in-vivo MPnRAGE data and T1 maps at 1.5 $\mathrm{mm}^3$ resolution. For our proposed INR reconstruction, slight slice artifacts are visible e.g. in the cortex for TI 1 as 3D volumes are obtained by stacking reconstructed 2D slices.
}}\label{fig:volunteer_retro_sag}
\end{figure*}
%

%
\begin{figure*}
\centerline{\includegraphics[width=\linewidth]{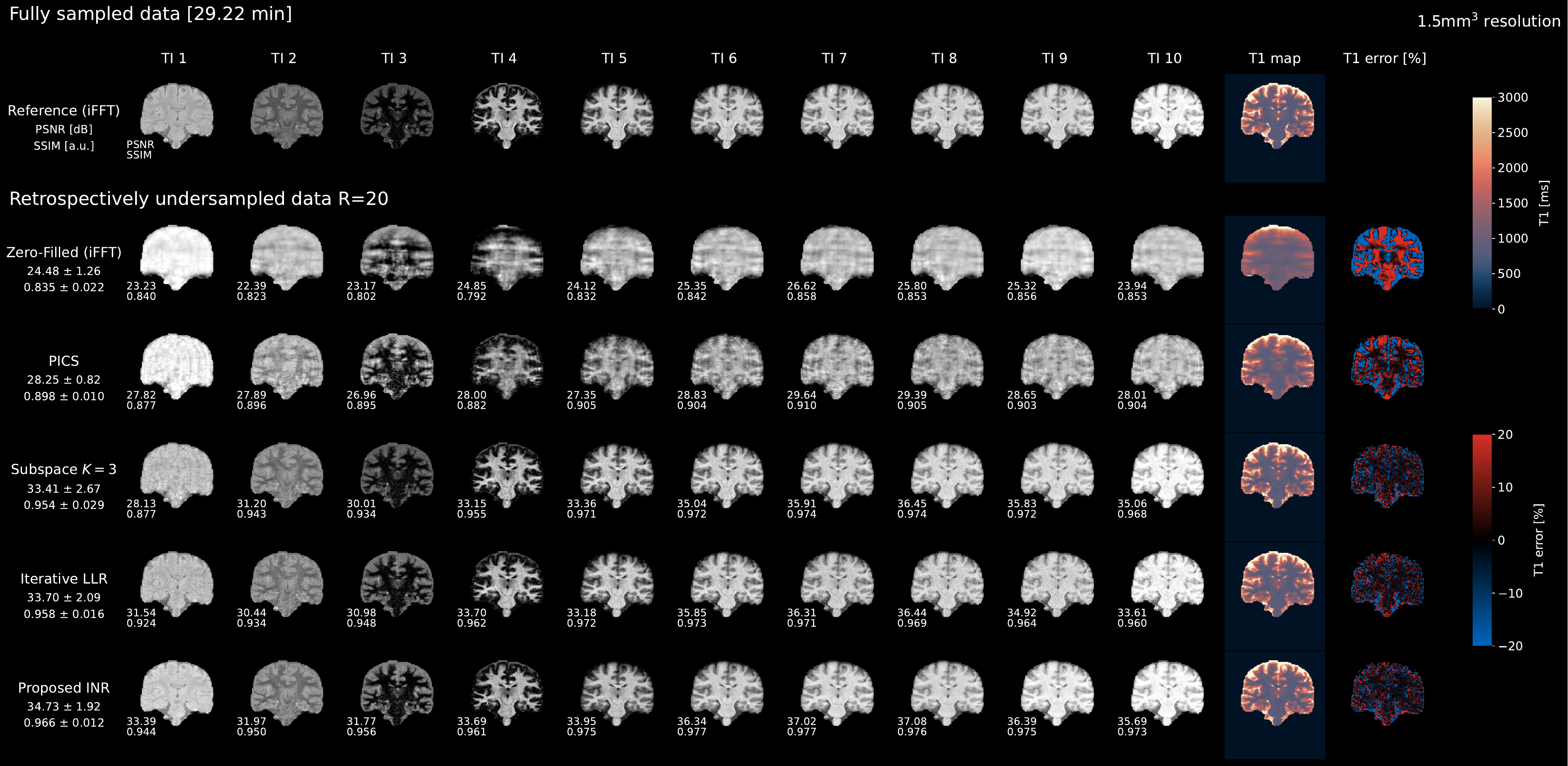}}
\caption{\hspace*{0.5em}{
Retrospective undersampling demonstrated for coronal in-vivo MPnRAGE data and T1 maps at 1.5 $\mathrm{mm}^3$ resolution.
}}\label{fig:volunteer_retro_cor}
\end{figure*}
%

%
\begin{figure*}
\centerline{\includegraphics[width=0.5\linewidth]{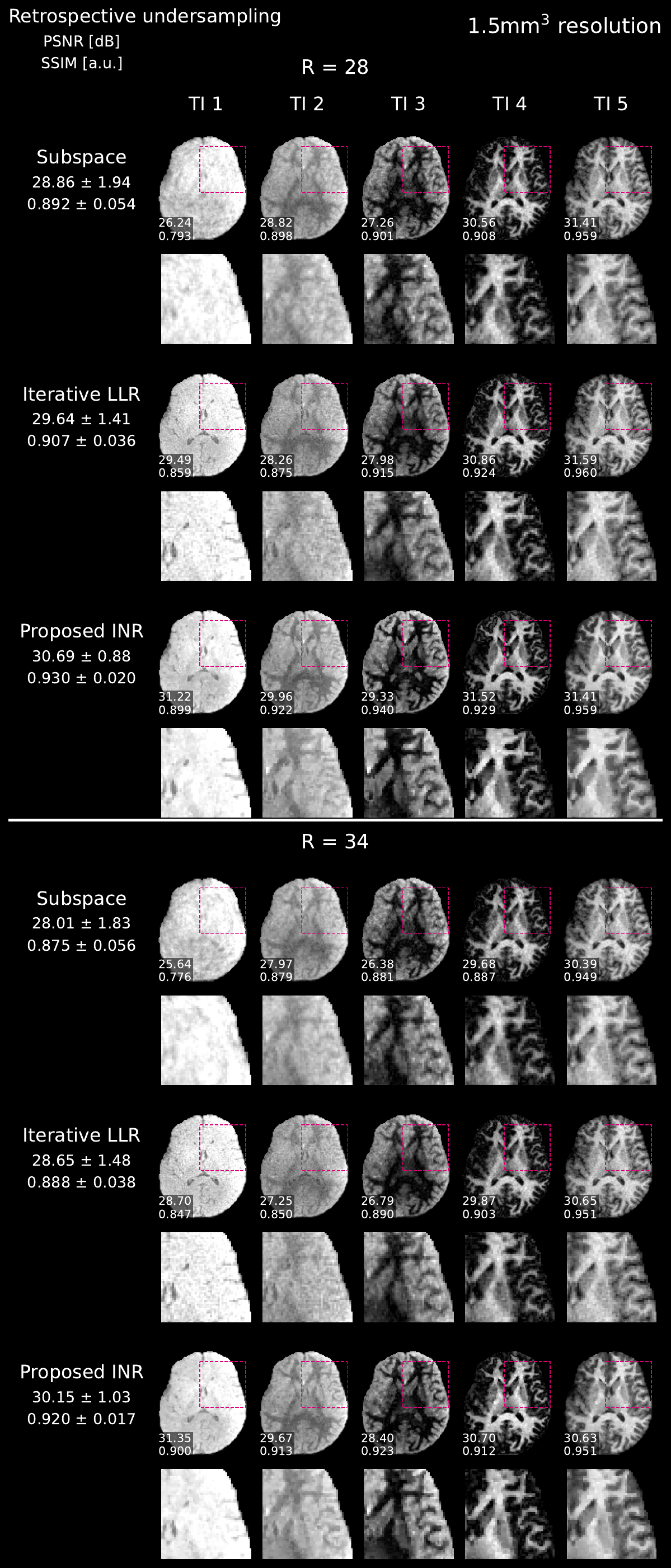}}
\caption{\hspace*{0.5em}{In-vivo MPnRAGE 1.5 $\mathrm{mm}^3$ resolution data retrospectively undersampled at acceleration factors R~=~28 \& 34 to assess the robustness of subspace, iterative LLR and INR reconstruction with increased undersampling. The image quality for subspace reconstruction and iterative LLR degrade more with increasing acceleration factor than for our proposed INR reconstruction, especially for the first three TI.
}}\label{fig:Supp_R_limit}
\end{figure*}
%

%
\begin{figure*}
\centerline{\includegraphics[width=\linewidth]{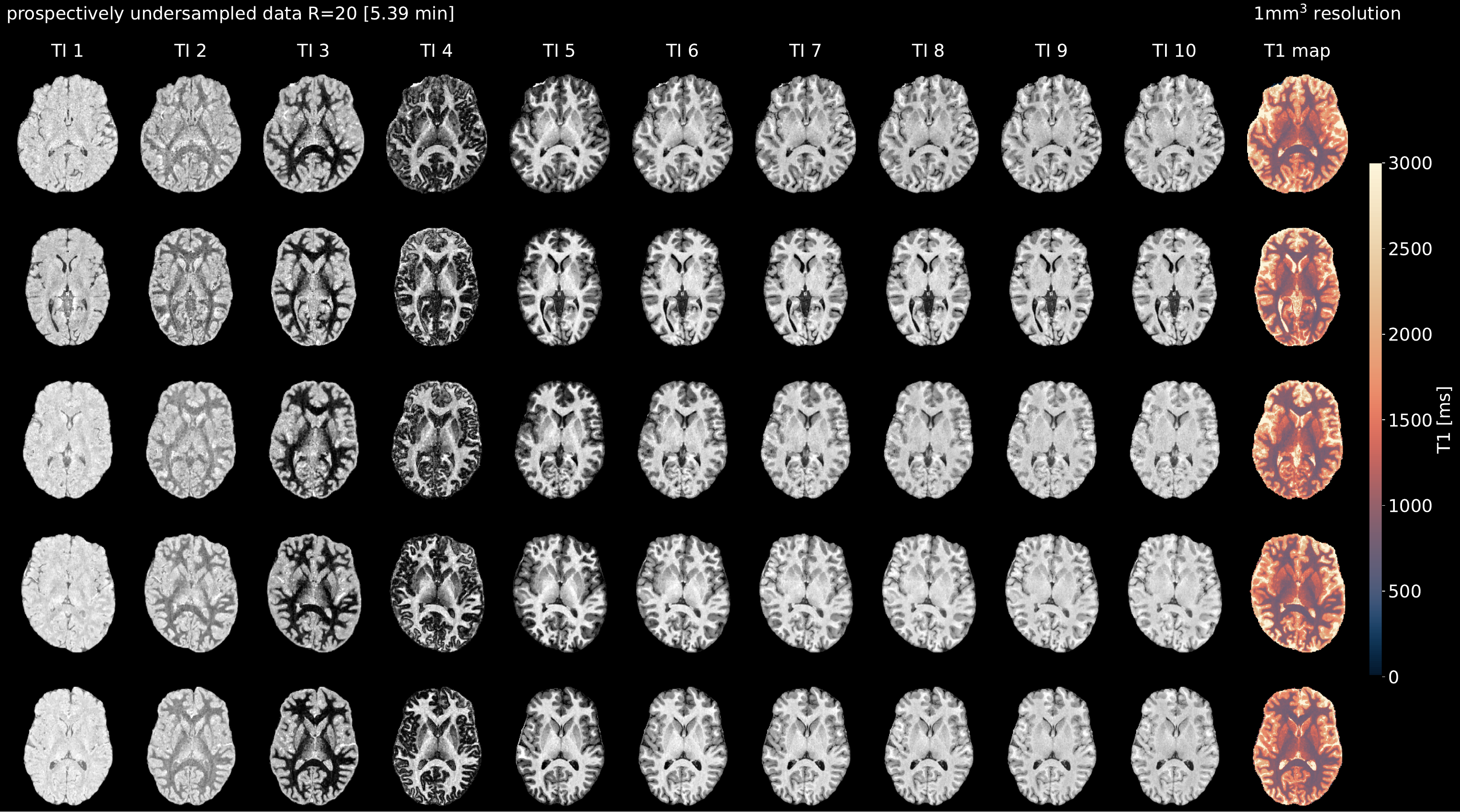}}
\caption{\hspace*{0.5em}{In-vivo MPnRAGE 1 $\mathrm{mm}^3$ resolution data of five different healthy volunteers acquired in 5.39 minutes (R~=~20), reconstructed with our proposed INR framework. Hyperparameters in this work are the same across all volunteers, resolutions and accelerations suggesting robustness across acquisition settings and generalizability.
}}\label{fig:Supp_all_volunteers}
\end{figure*}
%

%

\newcommand{\voltime}[1]{%
  \ifdim \fpeval{#1*240}pt < 3600pt
    \fpeval{round(#1*240/60, 1)} min%
  \else
    \fpeval{round(#1*240/3600, 2)} h%
  \fi
}

\newcommand{\reconrow}[2]{%
  #1 & #2 s & \voltime{#2} \\
}

\setcounter{table}{0}
\renewcommand{\thetable}{S\arabic{table}}

\begin{table*}[t] \centering \caption{\hspace*{0.5em}{Reconstruction times for 3D data prospectively accelerated at R~=~20, reconstructed on a NVIDIA RTX A2000 12 GB GPU.}} \label{tab:recon_time} \begin{tabular}{lcc} \toprule Method & Time per slice & Time per 3D volume \\ \midrule \reconrow{Subspace reconstruction}{1.64} \reconrow{Iterative LLR}{18.42} \reconrow{Proposed INR}{3.45} \bottomrule \end{tabular} \end{table*}

\end{document}